\documentclass[3p,12pt]{elsarticle}
\usepackage{amssymb,latexsym,amsmath,algorithm,algpseudocode}%hyperref,
\journal{}

\newcommand{\eps}{\varepsilon}

\newcommand{\set}[1]{\left\{#1\right\}}

\newcommand{\p}{\partial}

\newtheorem{thm}{Theorem}[section]

\newtheorem{Rem}[thm]{Remark}

\begin{document}
\begin{frontmatter}



\title{Multi-frequency based location search algorithm of small electromagnetic inhomogeneities embedded in two-layered medium}


\author{Won-Kwang Park}
\ead{parkwk@kookmin.ac.kr}
\author{Taehoon Park\corref{corPark}}
\ead{thpark@kookmin.ac.kr}
\cortext[corPark]{Corresponding author}

\address{Department of Mathematics, The College of Natural Sciences, Kookmin University, Seoul, 136-702, Korea.}

\begin{abstract}
In this paper, we consider a problem for finding the locations of electromagnetic inhomogeneities completely embedded in homogeneous two layered medium. For this purpose, we present a filter function operated at several frequencies and design an algorithm for finding the locations of such inhomogeneities. It is based on the fact that the collected Multi-Static Response (MSR) matrix can be modeled via a rigorous asymptotic expansion formula of the scattering amplitude due to the presence of such inhomogeneities. In order to show the effectiveness, we compare the proposed algorithm with traditional MUltiple SIgnal Classification (MUSIC) algorithm and Kirchhoff migration. Various numerical results demonstrate that the proposed algorithm is robust with respect to random noise and yields more accurate location than the MUSIC algorithm and Kirchhoff migration.
\end{abstract}


\end{frontmatter}





\section{Introduction}
In the non-destructive evaluation area, an inverse problem is finding the specific characteristics (location, geometry, internal constitution etc.) of small inhomogeneities from measurements of scattered or far-field data. This problem, which arises in fields such as physics, engineering, and biomedical science, is highly related to everyday human life and is still a challenging problem \cite{A,AK,DEKPS}.

Among them, finding the locations of anti-personnel mines (embedded in the soil) from electromagnetic data (measured in the air) is an interesting and important problem in the military services. The mines have different material properties from the surrounding medium and they are small relative to the area. The main purpose of this kind of application is to find their locations more accurately, not to retrieve complete information. For this purpose, non-iterative MUSIC-type algorithm for finding locations of small inhomogeneities buried within a half-space at a fixed frequency has been developed, refer to \cite{AIL}. MUSIC type algorithms are advantageous in the sense that they are fast, stable, can easily be extended to the multiple inhomogeneities, and they do not require specific regularization terms that are highly dependent on the problem at hand. However, due to the reason that one is faced with an aspect-limited inverse problem in the reflection mode because sources and receivers are located upper half-space, sometimes inexact locations are identified. Hence, in order to obtain an accurate locations of such inhomogeneities, an alternative algorithm that can overcome these problems is necessary.

Motivated by the above fact, we propose an effective, non-iterative location search algorithm at multiple frequencies that can work on limited view data in order to find the accurate locations of small electromagnetic inhomogeneities completely embedded within a homogeneous (lower) half-space. The starting point is that the collected MSR matrix can be approximated by a rigorous asymptotic expansion formula of the scattering amplitude in the presence of such inhomogeneities. This approximation leads us to proceed the singular value decomposition of the MSR matrix and to identify the structure of singular vectors. Applying the structure of singular vectors, a filter function operated at several frequencies can be designed for finding exact locations of inhomogeneities. Moreover, based on the statistical hypothesis testing, we can confirm that the proposed location search algorithm offers more exact information of locations than the established MUSIC-type algorithm and Kirchhoff migration.

Various numerical examples with noisy data will illustrate the behavior of the proposed location search algorithm operated at several frequencies. Unlike the numerical setting in \cite{AIL}, computational examples under the narrow/wide range of incident and observation directions configuration and the closely located small inclusions situation show the feasibilities and limitations of the proposed algorithm.

This paper is organized as follows. In section \ref{Sec2}, we briefly survey the two-dimensional direct scattering problem for two-layered medium and introduce an asymptotic expansion formula for the scattering amplitude. In section \ref{Sec3}, we design a location search algorithm by producing a filter function operated at several frequencies and compare its detection performance with traditional MUSIC-type algorithm and Kirchhoff migration. In section \ref{Sec4}, corresponding numerical experiments with random noise are shown and compared with the result via MUSIC-type one and Kirchhoff migration to demonstrate its performance. In section \ref{Sec5} we give a brief conclusion.

Finally, we will refer to \cite{G} as a useful reference that investigates the so-called reciprocity gap MUSIC as another algorithm linked to the MUSIC one. In this reference, the author shows that reconstructed positions of the small inhomogeneities via reciprocity gap MUSIC algorithm are slightly more accurate than the traditional MUSIC one.

\section{Survey on two-dimensional direct scattering problem}\label{Sec2}
In this section, we briefly discuss the two-dimensional, time-harmonic electromagnetic scattering from a small inhomogeneity buried within a homogeneous half-space. A more detailed description can be found in \cite{AIL}.

Let us decompose the two-dimensional space $\mathbb{R}^2$ into the lower and the upper half-spaces as
\[\mathbb{R}_{-}^{2}=\set{\mathbf{x}=(x_1,x_2)^T\in\mathbb{R}^2:x_2<0}\quad\mbox{and}\quad\mathbb{R}_{+}^{2}=\set{\mathbf{x}=(x_1,x_2)^T\in\mathbb{R}^2:x_2>0},\]
respectively and assume that these spaces are homogeneous. Let $\p\mathbb{R}^2$ denote the border between $\mathbb{R}_{-}^{2}$ and $\mathbb{R}_{+}^{2}$, represented as
\[\p\mathbb{R}^2=\set{(x_1,x_2)^T\in\mathbb{R}^2:x_2=0}.\]

Throughout this paper, we assume that all the electromagnetic small inhomogeneities $D_m$, $m=1,2,\cdots,M$, are completely embedded in the lower half-space $\mathbb{R}_{-}^{2}$. For convenience, let $\mathcal{D}$ denote the collection of such inhomogeneities as
\[\mathcal{D}=\bigcup_{m=1}^{M}D_m=\bigcup_{m=1}^{M}(\mathbf{z}_m+r B_m),\]
where $r$ is a small (with respect to the wavelength of the electromagnetic field in the embedding space at the given frequency of operation $\omega$) positive constant which denotes the diameter of the inhomogeneities, $B_j$ is a simply connected smooth domain containing the origin, and $\mathbf{z}_j\in\mathbb{R}_{-}^{2}$ indicates the location of inhomogeneities. Throughout this paper, we assume that these inhomogeneities are separated enough from each other and from $\p\mathbb{R}^2$.

All materials are characterized by their dielectric permittivity and magnetic permeability at the given frequency $\omega$; $\eps_-$, $\eps_+$ and $\eps_m$ denotes the electric permittivity of $\mathbb{R}_{-}^{2}$, $\mathbb{R}_{+}^{2}$ and $D_m$, respectively. The magnetic permeabilities $\mu_{-}$, $\mu_{+}$ and $\mu_{m}$ can be defined analogously. Using these notations, we can define the piecewise constant electric permittivity $0<\eps(\mathbf{x})<+\infty$ and magnetic permeability $0<\mu(\mathbf{x})<+\infty$ as
\[\eps(\mathbf{x})=\left\{\begin{array}{ccl}
\eps_{+}&\mbox{for}&\mathbf{x}\in\mathbb{R}_{+}^{2}\\
\eps_{-}&\mbox{for}&\mathbf{x}\in\mathbb{R}_{-}^{2}\backslash\overline{\mathcal{D}}\\
\eps_m&\mbox{for}&\mathbf{x}\in D_m
\end{array}\right.\quad\mbox{and}\quad
\mu(\mathbf{x})=\left\{\begin{array}{ccl}
\mu_{+}&\mbox{for}&\mathbf{x}\in\mathbb{R}_{+}^{2}\\
\mu_{-}&\mbox{for}&\mathbf{x}\in\mathbb{R}_{-}^{2}\backslash\overline{\mathcal{D}}\\
\mu_m&\mbox{for}&\mathbf{x}\in D_m,
\end{array}\right.\]
respectively. For convenience, we also define the electric permittivity $\eps_0(\mathbf{x})$ and magnetic permeability $\mu_0(\mathbf{x})$ when there are no inhomogeneities such as
\[\eps_0(\mathbf{x})=\left\{\begin{array}{ccl}
\eps_{+}&\mbox{for}&\mathbf{x}\in\mathbb{R}_{+}^{2}\\
\eps_{-}&\mbox{for}&\mathbf{x}\in\mathbb{R}_{-}^{2}
\end{array}\right.\quad\mbox{and}\quad
\mu_0(\mathbf{x})=\left\{\begin{array}{ccl}
\mu_{+}&\mbox{for}&\mathbf{x}\in\mathbb{R}_{+}^{2}\\
\mu_{-}&\mbox{for}&\mathbf{x}\in\mathbb{R}_{-}^{2},
\end{array}\right.\]
respectively. Accordingly, the piecewise positive real-valued wavenumber $k(\mathbf{x};\omega)$ reads as
\[k(\mathbf{x};\omega)=\left\{\begin{array}{ccl}
k_{+}(\omega)=\omega\sqrt{\eps_+\mu_+}&\mbox{for}&\mathbf{x}\in\mathbb{R}_{+}^{2}\\
k_{-}(\omega)=\omega\sqrt{\eps_-\mu_-}&\mbox{for}&\mathbf{x}\in\mathbb{R}_{-}^{2}.
\end{array}\right.\]

Let $\boldsymbol\theta=(\theta_1,\theta_2)^T$ be a two-dimensional vector on the unit circle $S^1\subset\mathbb{R}^2$ and $u_{\mathrm{inc}}(\mathbf{x})=\exp(ik_+(\omega)\boldsymbol\theta\cdot\mathbf{x})$ be a planar incident wavefield generated in the upper half-space on $\mathbb{R}_{+}^{2}$. At a given frequency $\omega$, $u_{\mathrm{tot}}(\mathbf{x};\omega)$ denotes the time-harmonic electromagnetic total field which satisfies the following two-dimensional Helmholtz equation
\begin{equation}\label{Helmholtz}
  \nabla\cdot\left(\frac{1}{\mu(\mathbf{x})}\nabla u_{\mathrm{tot}}(\mathbf{x};\omega)\right)+\omega^2 \eps(\mathbf{x})u_{\mathrm{tot}}(\mathbf{x};\omega)=0\quad\mbox{in}\quad\mathbb{R}^2,
\end{equation}
with transmission conditions holding at boundaries $\p\mathbb{R}^2$ and $\p D_m$ as
\[[u_{\mathrm{tot}}(\mathbf{x};\omega)]=\left[\frac{1}{\mu(\mathbf{x})}\frac{\p u_{\mathrm{tot}}(\mathbf{x};\omega)}{\p x_2}\right]=0~~\mbox{on}~~\p\mathbb{R}^2~~\mbox{and}~~[u_{\mathrm{tot}}(\mathbf{x};\omega)]=\left[\frac{1}{\mu(\mathbf{x})}\frac{\p u_{\mathrm{tot}}(\mathbf{x};\omega)}{\p\mathbf{\nu}_m(\mathbf{x})}\right]=0~~\mbox{on}~~\p D_m,\]
respectively. Here, $\mathbf{\nu}_m$ denotes the unit outward normal to $\p D_m$, $m=1,2,\cdots,m$.

%Then, within each half-space, scattered field $u_{\mathrm{scat}}(\mathbf{x};\omega)=u_{\mathrm{tot}}(\mathbf{x};\omega)-u_{\mathrm{bac}}(\mathbf{x};\omega)$ satisfies the Sommerfeld radiation condition
%\[\lim_{|\mathbf{x}|\to\infty}\sqrt{|\mathbf{x}|}\left(\frac{\p u_{\mathrm{scat}}(\mathbf{x};\omega)}{\p|\mathbf{x}|}-ik(\mathbf{x})u_{\mathrm{scat}}(\mathbf{x};\omega)\right)=0\]
%uniformly in all directions $\frac{\mathbf{x}}{|\mathbf{x}|}$.

Let $u_{\mathrm{bac}}(\mathbf{x};\omega)$ be the solution to the Helmholtz equation (\ref{Helmholtz}) in the absence of inhomogeneities. Then the scattering amplitude is defined as a function $K:(S^1\backslash\{\mathbf{0}\})\times(S^1\backslash\{\mathbf{0}\})\times\mathbb{R}\longrightarrow\mathbb{C}$ that satisfies
\[u_{\mathrm{tot}}(\mathbf{x};\omega)-u_{\mathrm{bac}}(\mathbf{x};\omega)=\frac{\{k_-(\omega)\}^2\mu_+(1+i)}{4\mu_-\sqrt{k_+(\omega)\pi}} \frac{\exp(ik(\mathbf{y};\omega)|\mathbf{y}|)}{\sqrt{|\mathbf{y}|}}K(\boldsymbol\vartheta,\boldsymbol\theta;\omega)+o\left(\frac{1}{\sqrt{|\mathbf{y}|}}\right)\]
as $|\mathbf{y}|\longrightarrow\infty$ uniformly on $\boldsymbol\vartheta=\frac{\mathbf{y}}{|\mathbf{y}|}$.

In order to represent asymptotic expansion formula of $K(\boldsymbol\vartheta,\boldsymbol\theta)$, we need some ingredients. First, based on the fact that the scattered field data acquisition is possible only on the upper half-space $\mathbb{R}_{+}^2$, we divide the unit circle $S^1$ into
\[S_{-}^1=\mathbb{R}_{-}^2\cap S^1\quad\mbox{and}\quad S_{+}^1=\mathbb{R}_{+}^2\cap S^1.\]
Next, by letting $\xi=\frac{k_+(\omega)}{k_-(\omega)}$, define a vector $\boldsymbol\phi(\boldsymbol\theta;\omega):S^1\backslash\{\mathbf{0}\}\longrightarrow\mathbb{C}^{2\times1}$ as
\begin{equation}\label{vecphi}
  \boldsymbol\phi(\boldsymbol\theta;\omega)=\left(\xi\theta_1,\mbox{sign}(\theta_2)\sqrt{1-\xi^2\theta_1^2}\right)^T,
\end{equation}
and a function $\Phi(\boldsymbol\theta;\omega):S^1\backslash\{\mathbf{0}\}\longrightarrow\mathbb{C}$ as
\begin{equation}\label{functionPhi}
  \Phi(\boldsymbol\theta)=\frac{2\mu_-\xi\theta_2}{\mu_-\xi\theta_2+\mu_+\mbox{sign}(\theta_2)\sqrt{1-\xi^2\theta_1^2}},
\end{equation}
respectively. With these, an asymptotic expansion formula of scattering amplitude can be written as follows. This formula plays a key role of the location search algorithm that will be designed in the next section.

\begin{thm}\label{TheoremAsymptotic}For every $\boldsymbol\vartheta\in S_{+}^1$ and $\boldsymbol\theta\in S_{-}^1$, the asymptotic formula for the scattering amplitude $K(\boldsymbol\vartheta,\boldsymbol\theta)$ at frequency $\omega$ is expressed as
\begin{align}
\begin{aligned}\label{SAK}
  K(\boldsymbol\vartheta,\boldsymbol\theta;\omega)=&r^2\Phi(\boldsymbol\vartheta;\omega)\Phi(\boldsymbol\theta;\omega)\sum_{m=1}^{M}\bigg(\gamma_{\eps,m}|B_m|+\gamma_{\mu,m}\boldsymbol\phi(\boldsymbol\vartheta;\omega)\cdot\mathbb{P}(\mathbf{z}_m)\cdot\boldsymbol\phi(\boldsymbol\theta;\omega)\bigg)\times\\
  &\exp\bigg(-ik_-(\omega)[\boldsymbol\phi(\boldsymbol\vartheta;\omega)-\boldsymbol\phi(\boldsymbol\theta;\omega)]\cdot \mathbf{z}_m\bigg)+o(r^2),
  \end{aligned}
\end{align}
where the remaining term $o(r^2)$ is independent of $\boldsymbol\vartheta\in S_+^1$, $\boldsymbol\theta\in S_-^1$ and the set of points $\set{\mathbf{z}_m}_{m=1}^{M}$,  $\mathbb{P}(\mathbf{z}_m)$ is a $2\times2$ positive, symmetric matrix
\[\mathbb{P}(\mathbf{z}_m)=\frac{2\mu_-}{\mu_-+\mu_m}|B_m|\mathbb{I}_2,\]
and constants $\gamma_{\mu,m}$ and $\gamma_{\eps,m}$ are given by \[\gamma_{\eps,m}=\frac{\eps_m}{\eps_-}-1\quad\mbox{and}\quad\gamma_{\mu,m}=\frac{\mu_m}{\mu_-}-1.\]
\end{thm}

\section{Non-iterative location search algorithm \& its performance}\label{Sec3}
\subsection{Non-iterative location search algorithm at multiple frequencies}
We apply the asymptotic formula for the scattering amplitude (\ref{SAK}) in order to build up the location search algorithm. For this purpose, we will use the eigenvalue structure of the Multi-Static Response (MSR) matrix $\mathbb{K}:=\mathbb{K}(\omega)=(K_{jl}(\omega))\in\mathbb{C}^{N_{\mbox{\tiny obs}}\times N_{\mbox{\tiny inc}}}$, whose element $K_{jl}(\omega):=K(\boldsymbol\vartheta_j,\boldsymbol\theta_l;\omega)$ is the scattering amplitude collected at observation number $j$ for the incident wave numbered $l$. In this paper, we denote
\[\{\boldsymbol\vartheta_j:j=1,2,\cdots,N_{\mbox{\tiny obs}}\}\quad\mbox{and}\quad \{\boldsymbol\theta_l:l=1,2,\cdots,N_{\mbox{\tiny inc}}\}\]
be the set of observation and incident directions, respectively. Note that when the upper half-space is more refractive than the lower one, i.e., if $k_+>k_-$, the number $N_{\mbox{\tiny obs}}$ of propagating transmitted waves might be less than $N_{\mbox{\tiny inc}}$ (see \cite{AIL} for instance). Then since $jl-$th element of the MSR matrix $K_{jl}(\omega)$ can be approximated as
\begin{align}
\begin{aligned}\label{MSR}
K_{jl}(\omega)=&K(\boldsymbol\vartheta_j,\boldsymbol\theta_l;\omega)\\
\approx&r^2\Phi(\boldsymbol\vartheta_j;\omega)\Phi(\boldsymbol\theta_l;\omega)\sum_{m=1}^{M}\bigg(\gamma_{\eps,m}|B_m|+\gamma_{\mu,m}\boldsymbol\phi(\boldsymbol\vartheta;\omega)\cdot\mathbb{P}(\mathbf{z}_m)\cdot\boldsymbol\phi(\boldsymbol\theta;\omega)\bigg)\times\\
&\exp\bigg(-ik_-(\omega)[\boldsymbol\phi(\boldsymbol\vartheta_j;\omega)-\boldsymbol\phi(\boldsymbol\theta_l;\omega)]\cdot \mathbf{z}_m\bigg),
\end{aligned}
\end{align}
MSR matrix $\mathbb{K}$ can be decomposed as follows:
\begin{equation}\label{EVD}
\mathbb{K}=\mathbb{DEH}^T,
\end{equation}
where $\mathbb{E}\in\mathbb{C}^{3M\times3M}$ is a diagonal matrix with components
\[\mathbb{E}=\left[\begin{array}{cc}\mathbb{E}_{\eps}&0\\0&\mathbb{E}_{\mu}\end{array}\right]\]
for
\begin{align*}
\mathbb{E}_{\eps}&=M\times M\mbox{ diagonal matrix with components }r^2\gamma_{\eps,m}|B_m|,\\
\mathbb{E}_{\mu}&=2M\times2M\mbox{ diagonal matrix with }2\times2\mbox{ blocks }-r^2\gamma_{\mu,m}\mathbb{P}(\mathbf{z}_m),
\end{align*}
matrix $\mathbb{D}\in\mathbb{C}^{N_{\mbox{\tiny obs}}\times3M}$ is of the form
\[\left[\mathbb{D}_\eps^1\quad\mathbb{D}_\eps^2\quad\cdots\quad\mathbb{D}_\eps^M\quad\mathbb{D}_\mu^1\quad\mathbb{D}_\mu^2\quad\cdots\quad \mathbb{D}_\mu^{2M}\right]\]
with $N_{\mbox{\tiny obs}}\times 1$ matrices
\begin{align*}
\mathbb{D}_{\eps}^m=\bigg[&\Phi(\boldsymbol\vartheta_1;\omega)\exp\bigg(-ik_-(\omega)\boldsymbol\phi(\boldsymbol\vartheta_1;\omega)\cdot\mathbf{z}_m\bigg),\cdots,\\
&\Phi(\boldsymbol\vartheta_{N_{\mbox{\tiny obs}}};\omega)\exp\bigg(-ik_-(\omega)\boldsymbol\phi(\boldsymbol\vartheta_{N_{\mbox{\tiny obs}}};\omega)\cdot\mathbf{z}_m\bigg)\bigg]^T,\\
\mathbb{D}_{\mu}^{2(m-1)+s}=\bigg[&\mathbf{e}_s\cdot \boldsymbol\phi(\boldsymbol\vartheta_1;\omega)\Phi(\boldsymbol\vartheta_1;\omega)\exp\bigg(-ik_-(\omega) \boldsymbol\phi(\boldsymbol\vartheta_1;\omega)\cdot\mathbf{z}_m\bigg),\cdots,\\
&\mathbf{e}_s\cdot\boldsymbol\phi(\boldsymbol\vartheta_{N_{\mbox{\tiny obs}}};\omega)\Phi(\boldsymbol\vartheta_{N_{\mbox{\tiny obs}}};\omega)\exp\bigg(-ik_-(\omega)\boldsymbol\phi(\boldsymbol\vartheta_{N_{\mbox{\tiny obs}}};\omega)\cdot\mathbf{z}_m\bigg)\bigg]^T,
\end{align*}
and matrix $\mathbb{H}\in\mathbb{C}^{N_{\mbox{\tiny inc}}\times3M}$ can be written as follows
\[\left[\mathbb{H}_\eps^1\quad\mathbb{H}_\eps^2\quad\cdots\quad\mathbb{H}_\eps^M\quad\mathbb{H}_\mu^1\quad\mathbb{H}_\mu^2\quad\cdots\quad \mathbb{H}_\mu^{2M}\right]\]
with $N_{\mbox{\tiny inc}}\times 1$ matrices
\begin{align*}
\mathbb{H}_{\eps}^m=\bigg[&\Phi(\boldsymbol\theta_1;\omega) \exp\bigg(ik_-(\omega)\boldsymbol\phi(\boldsymbol\theta_1;\omega)\cdot\mathbf{z}_m\bigg),\cdots, \Phi(\boldsymbol\theta_{N_{\mbox{\tiny inc}}};\omega)\exp\bigg(ik_-(\omega)\boldsymbol\phi(\boldsymbol\theta_{N_{\mbox{\tiny inc}}};\omega)\cdot \mathbf{z}_m\bigg)\bigg]^T,\\
\mathbb{H}_{\mu}^{2(m-1)+s}=\bigg[&\mathbf{e}_s\cdot \boldsymbol\phi(\boldsymbol\theta_1;\omega)\Phi(\boldsymbol\theta_1;\omega)\exp\bigg(ik_-(\omega) \boldsymbol\phi(\boldsymbol\theta_1;\omega)\cdot\mathbf{z}_m\bigg),\cdots,\\
&\mathbf{e}_s\cdot\boldsymbol\phi(\boldsymbol\theta_{N_{\mbox{\tiny inc}}};\omega)\Phi(\boldsymbol\theta_{N_{\mbox{\tiny inc}}};\omega)\exp\bigg(ik_-(\omega)\boldsymbol\phi(\boldsymbol\vartheta_{N_{\mbox{\tiny inc}}};\omega)\cdot\mathbf{z}_m\bigg)\bigg]^T.
\end{align*}
Here $\set{\mathbf{e}_s}_{s=1,2}=\set{\mathbf{e}_1=(1,0)^T,\mathbf{e}_2=(0,1)^T}$ is an orthonormal basis of $\mathbb{R}^2$. Based on the decomposition (\ref{EVD}), a location search algorithm can be established as follows.

\begin{enumerate}
  \item (Singular Value Decomposition) Let us perform Singular Value Decomposition (SVD) of matrix $\mathbb{K}$ and let $M$ be the number of nonzero singular values for the given $\omega$. Then, $\mathbb{K}$ can be represented as follows:
      \[\mathbb{K}=\mathbb{U}(\omega)\mathbb{S}(\omega)\mathbb{V}^*(\omega)\approx\sum_{m=1}^{M}\sigma_m(\omega)\mathbf{U}_m(\omega)\mathbf{V}_m^*(\omega),\]
      where superscript $*$ denotes the complex conjugate, $\sigma_m(\omega)$ are the singular values, $\mathbf{U}_m(\omega)$ and $\mathbf{V}_m(\omega)$ are the left and right
      singular vectors of $\mathbb{K}$, respectively for $m=1,2,\cdots,M$.
  \item (Structure of singular vectors) For test vectors $\mathbf{c_d},\mathbf{c_h}\in\mathbb{R}^3\backslash\set{0}$, define vectors\\ $\mathbf{d}:\mathbb{R}_+^2\times\mathbb{R}\longrightarrow\mathbb{C}^{N_{\mbox{\tiny obs}}\times 1}$ and $\mathbf{h}:\mathbb{R}_-^2\times\mathbb{R}\longrightarrow\mathbb{C}^{N_{\mbox{\tiny inc}}\times 1}$as
\begin{align}
\begin{aligned}\label{VecD}
  \mathbf{d}(\mathbf{x};\omega)=\bigg[&\mathbf{c_d}\cdot(1,\boldsymbol\phi(\boldsymbol\vartheta_1;\omega))\Phi(\boldsymbol\vartheta_1;\omega)\exp\bigg(-ik_-(\omega)\boldsymbol\phi(\boldsymbol\vartheta_1;\omega)\cdot\mathbf{x}\bigg),\cdots,\\ &\mathbf{c_d}\cdot(1,\boldsymbol\phi(\boldsymbol\vartheta_{N_{\mbox{\tiny obs}}};\omega))\Phi(\boldsymbol\vartheta_{N_{\mbox{\tiny obs}}};\omega)\exp\bigg(-ik_-(\omega)\boldsymbol\phi(\boldsymbol\vartheta_{N_{\mbox{\tiny obs}}};\omega)\cdot\mathbf{x}\bigg)\bigg]^T
\end{aligned}
\end{align}
and
\begin{align}
\begin{aligned}\label{VecH}
  \mathbf{h}(\mathbf{x};\omega)=\bigg[&\mathbf{c_h}\cdot(1,\boldsymbol\phi(\boldsymbol\theta_1;\omega))\Phi(\boldsymbol\theta_1;\omega)\exp\bigg(ik_-(\omega)\boldsymbol\phi(\boldsymbol\theta_1;\omega)\cdot\mathbf{x}\bigg),\cdots,\\ &\mathbf{c_h}\cdot(1,\boldsymbol\phi(\boldsymbol\theta_{N_{\mbox{\tiny inc}}};\omega))\Phi(\boldsymbol\theta_{N_{\mbox{\tiny inc}}};\omega)\exp\bigg(ik_-(\omega)\boldsymbol\phi(\boldsymbol\theta_{N_{\mbox{\tiny inc}}};\omega)\cdot\mathbf{x}\bigg)\bigg]^T,
\end{aligned}
\end{align}
respectively. With this, generate corresponding normalized unit vectors
\[\mathbf{W_D}(\mathbf{x};\omega)=\frac{\mathbf{d}(\mathbf{x};\omega)}{|\mathbf{d}(\mathbf{x};\omega)|} \quad\mbox{and}\quad \mathbf{W_H}(\mathbf{x};\omega)=\frac{\mathbf{h}(\mathbf{x};\omega)}{|\mathbf{h}(\mathbf{x};\omega)|}.\]
Note that the structure of vectors $\mathbf{W_D}(\mathbf{x};\omega)$ and $\mathbf{W_H}(\mathbf{x};\omega)$ is motivated by the matrix $\mathbb{D}$ and $\mathbb{H}$ in (\ref{EVD}), respectively. Then by virtue of \cite{HHSZ}, following relationship holds for $m=1,2,\cdots,M$,
\begin{equation}\label{SimilarVector}
  \mathbf{U}_m(\omega)\sim \mathbf{W_D}(\mathbf{z}_m;\omega)\quad\mbox{and}\quad\overline{\mathbf{V}}_m(\omega)\sim \mathbf{W_H}(\mathbf{z}_m;\omega).
\end{equation}
Since the first $M$ columns of the matrix $\mathbb{U}(\omega)$ and $\mathbb{V}(\omega)$, $\set{\mathbf{U}_1(\omega),\mathbf{U}_2(\omega),\cdots,\mathbf{U}_m(\omega)}$ and $\set{\mathbf{V}_1(\omega),\mathbf{V}_2(\omega),\cdots,\mathbf{V}_m(\omega)}$, are orthonormal, we can observe that
\begin{align}
\begin{aligned}\label{ApproxSV}
  &\mathbf{W_D}(\mathbf{x};\omega)^*\mathbf{U}_m(\omega)\ne0\quad\mbox{and}\quad\mathbf{W_H}(\mathbf{x};\omega)^*\overline{\mathbf{V}}_m(\omega)\ne0,\quad\mbox{if}\quad\mathbf{x}=\mathbf{z}_m\\
  &\mathbf{W_D}(\mathbf{x};\omega)^*\mathbf{U}_m(\omega)\approx0\quad\mbox{and}\quad\mathbf{W_H}(\mathbf{x};\omega)^*\overline{\mathbf{V}}_m(\omega)\approx0,\quad\mbox{if}\quad\mathbf{x}\ne\mathbf{z}_m,
\end{aligned}
\end{align}
for $m=1,2,\cdots,M$.
  \item (Filter function) For a search domain $\Omega\subset\mathbb{R}_-^2$, construct a normalized filter function $\mathbb{F}:\Omega\times\mathbb{N}\longrightarrow\mathbb{R}$ at several frequencies $\{\omega_f:f=1,2,\cdots,F\}$ as
\begin{equation}\label{Imagefunctionmultiple}
\mathbb{F}(\mathbf{x};F)=\frac{1}{F}\left|\sum_{f=1}^{F}\sum_{m=1}^{M}\bigg(\mathbf{W_D}(\mathbf{x};\omega_f)^*\mathbf{U}_m(\omega_f)\bigg) \bigg(\mathbf{W_H}(\mathbf{x};\omega_f)^*\overline{\mathbf{V}}_m(\omega_f)\bigg)\right|.
\end{equation}
Then, based on the observation (\ref{ApproxSV}), we can find locations $\mathbf{z}_m\in D_m$ by finding $\mathbf{x}$ which satisfies $\mathbb{F}(\mathbf{x};F)\approx1$.
\end{enumerate}

\begin{Rem}[A priori information]
  In order to build up the filter (\ref{Imagefunctionmultiple}), we need a priori information of the values $\eps_-$ and $\mu_-$. If one has no information of $\eps_-$ or $\mu_-$, simultaneous reconstruction of multiple parameters must be performed (see \cite[Section 10.3]{DL} for instance).
\end{Rem}

The location search algorithm is summarized in Algorithm \ref{LSA}.

\begin{algorithm}
\begin{algorithmic}[1]
\Procedure{LSA}{$F$}%\Comment{Location Search Algorithm with $F$-frequencies}
\State identify values $\eps_-$ and $\mu_-$
\State initialize $K(\mathbf{x})$
\For{$f=1$ \textbf{to} $F$}
   \State collect MSR matrix data $\mathbb{K}(\omega_f)$
   \State perform SVD of $\mathbb{K}(\omega_f)$
   \State discriminate nonzero singular values%\Comment{see \cite[Section 4.2.1]{PL3}}
   \State choose $\mathbf{U}_m(\omega_f)$ and $\mathbf{V}_m(\omega_f)$
   \For{$\mathbf{x}\in\Omega\subset\mathbb{R}_{-}^{2}$}%\Comment{$\Omega$ is a search domain}
      \State generate $\mathbf{W_D}(\mathbf{x};\omega_f)$ and $\mathbf{W_H}(\mathbf{x};\omega_f)$%\Comment{based on (\ref{VecD})}
      \State initialize $I(\mathbf{x},f)$
         \For{$m=1$ \textbf{to} $M$}
            \State $I(\mathbf{x},f)\gets I(\mathbf{x},f)+(\mathbf{W_D}(\mathbf{x};\omega_f)^*\mathbf{U}_m(\omega_f))(\mathbf{W_H}(\mathbf{x};\omega_f)^*\overline{\mathbf{V}}_m(\omega_f))$.
         \EndFor
      \State $K(\mathbf{x})\gets I(\mathbf{x},f)$
   \EndFor
\EndFor
\State plot $\mathbf{F}(\mathbf{x};F)=|K(\mathbf{x})|/F$
\State find $\mathbf{x}=\mathbf{z}_m\in D_m$%\Comment{$\mathbf{F}(\mathbf{x};F)\approx1$}
\EndProcedure
\end{algorithmic}
\caption{Location Search Algorithm}\label{LSA}
\end{algorithm}

\subsection{Some properties of normalized filter function $\mathbb{F}$}\label{Sec3sub}
At this moment, we explore some properties of normalized filter function $\mathbb{F}$ in (\ref{Imagefunctionmultiple}). For this purpose, we assume that there exists one inhomogeneity ($m=1$) and
\begin{equation}\label{SelectionC}
  \mathbf{c_d}\cdot(1,\boldsymbol\phi(\boldsymbol\theta_n;\omega_f))\ne0,\quad \mathbf{c_h}\cdot(1,\boldsymbol\phi(\boldsymbol\theta_n;\omega_f))\ne0\quad\mbox{and}\quad\Phi(\boldsymbol\theta_n;\omega_f)\ne0
\end{equation}
for all $n=1,2,\cdots,N$ and $f=1,2,\cdots,F$. Then applying relation (\ref{SimilarVector}) to (\ref{Imagefunctionmultiple}) yields
\begin{align*}
  \mathbb{F}(\mathbf{x};F)&=\frac{1}{F}\left|\sum_{f=1}^{F}\bigg(\mathbf{W_D}(\mathbf{x};\omega_f)^*\mathbf{U}_1(\omega_f)\bigg) \bigg(\mathbf{W_H}(\mathbf{x};\omega_f)^*\overline{\mathbf{V}}_1(\omega_f)\bigg)\right| =\frac{1}{F}\left|\mathbb{F}_{\mathbf{D}}(\mathbf{x};F)\mathbb{F}_{\mathbf{H}}(\mathbf{x};F)\right|,
\end{align*}
where
\begin{align*}
  \mathbb{F}_{\mathbf{D}}(\mathbf{x};F)&=\sum_{n=1}^{N}\frac{|\mathbf{c_d}\cdot(1,\boldsymbol\phi(\boldsymbol\vartheta_n;\omega_f))\Phi(\boldsymbol\vartheta_n;\omega_f)|^2} {|\mathbf{d}(\mathbf{z}_m;\omega_f)||\mathbf{d}(\mathbf{x};\omega_f)|} \exp\bigg(-ik_-(\omega_f)\boldsymbol\phi(\boldsymbol\vartheta_n;\omega_f)\cdot(\mathbf{z}_m-\mathbf{x})\bigg)\\
  \mathbb{F}_{\mathbf{H}}(\mathbf{x};F)&=\sum_{n=1}^{N}\frac{|\mathbf{c_h}\cdot(1,\boldsymbol\phi(\boldsymbol\theta_n;\omega_f))\Phi(\boldsymbol\theta_n;\omega_f)|^2} {|\mathbf{h}(\mathbf{z}_m;\omega_f)||\mathbf{h}(\mathbf{x};\omega_f)|} \exp\bigg(ik_-(\omega_f)\boldsymbol\phi(\boldsymbol\theta_n;\omega_f)\cdot(\mathbf{z}_m-\mathbf{x})\bigg).
\end{align*}
With this, by letting $\boldsymbol\vartheta_n:=(\vartheta_1^n,\vartheta_2^n)^T$, $\boldsymbol\theta_n:=(\theta_1^n,\theta_2^n)^T$ and $\mathbf{z}:=(z_1,z_2)^T$, we can observe that
\begin{align*}
  \mathbb{F}_{\mathbf{D}}(\mathbf{x};F)&\sim\sum_{n=1}^{N}\exp\bigg(-ik_-(\omega_f)\left(\xi\vartheta_1^n(z_1-x_1) +\sqrt{1-(\xi\vartheta_1^n)^2}(z_2-x_2)\right)\bigg)\\
  \mathbb{F}_{\mathbf{H}}(\mathbf{x};F)&\sim\sum_{n=1}^{N}\exp\bigg(ik_-(\omega_f)\left(\xi\theta_1^n(z_1-x_1) -\sqrt{1-(\xi\theta_1^n)^2}(z_2-x_2)\right)\bigg)
\end{align*}
where $A\sim B$ means that there exists a constant $C$ such that $A=BC$. Throughout this paper, we assume that $N_{\mbox{\tiny obs}}$ and $N_{\mbox{\tiny inc}}$ are even number $N_{\mbox{\tiny obs}}=2L$, $\{\boldsymbol\vartheta_n\}$ and $\{\boldsymbol\theta_n\}$ are symmetric to $y-$axis, i.e., if $\boldsymbol\vartheta_n:=(\vartheta_1^n,\vartheta_2^n)^T$ then $\vartheta_1^n=-\vartheta_1^{2L-n-1}$ and $\vartheta_2^n=\vartheta_2^{2L-n-1}$, and so on. Then we can explore some properties of $\mathbb{F}(\mathbf{x};F)$ as follows

\begin{enumerate}
  \item Assume that $k_+(\omega_f)$ satisfies $k_+(\omega_f)>k_-(\omega_f)$ and $\xi\vartheta_1^n>1$. Then since $\sqrt{1-(\xi\vartheta_1^n)^2}\in\mathbb{C}\backslash\{0\}$, by letting $\sqrt{1-(\xi\vartheta_1^n)^2}=i\rho_n$, $\mathbb{F}_{\mathbf{D}}(\mathbf{x};F)$ can be written as
      \[\mathbb{F}_{\mathbf{D}}(\mathbf{x};F)\sim\sum_{n=1}^{N}\exp\bigg(-ik_+(\omega_f)\xi\vartheta_1^n(z_1-x_1)\bigg) \exp\bigg(\rho_n k_-(\omega_f)(z_2-x_2)\bigg).\]
      This means that $\mathbb{F}(\mathbf{x};F)\approx1$ when \[x_1=z_1+\frac{s\pi}{k_+(\omega_f)\theta_1^n},\quad x_1=z_1+\frac{s\pi}{k_+(\omega_f)\theta_1^n}+\frac{\pi}{2}\quad\mbox{and}\quad x_2=z_2\]
      for all $s\in\mathbb{Z}$, $f=1,2,\cdots,F$, and $n=1,2,\cdots,N_{\mbox{\tiny obs}}$. This relation tells us that in order to obtain an accurate location $\mathbf{z}$, one must applies high frequency and large number $N_{\mbox{\tiny obs}}$. Nevertheless, some replicas will appear along the $x-$axis. However, if one adopt symmetric observation configuration, $\mathbb{F}_{\mathbf{D}}(\mathbf{x};F)$ becomes
      \[\mathbb{F}_{\mathbf{D}}(\mathbf{x};F)\sim\sum_{n=1}^{L}\cos\bigg(-ik_+(\omega_f)\xi\vartheta_1^n(z_1-x_1)\bigg) \exp\bigg(\rho_n k_-(\omega_f)(z_2-x_2)\bigg).\]
      Hence, we can obtain more accurate location $\mathbf{z}$. See Figure \ref{Symmetric}.
  \item When the value $F$ is large enough, due to the different values of $\omega_f$, $\mathbb{F}(\mathbf{x};F)$ will yields more accurate location $\mathbf{z}$. This means that application of multiple frequencies will enhance detection performance.
  \item Assume that $k_+(\omega_f)\approx k_-(\omega_f)$ and $\vartheta_1^n\longrightarrow\pm1$ for some $n$, i.e., one has wide observation direction, then since $\sqrt{1-(\xi\vartheta_1^n)^2}\approx0$, location $\mathbf{x}$ such that $\mathbb{F}(\mathbf{x};F)\approx1$ is independent to the $z_2$. Hence we cannot identify location $\mathbf{z}$ via $\mathbb{F}(\mathbf{x};F)$, refer to Figure \ref{InfluenceRange}.
  \item If $k_+(\omega_f)<k_-(\omega_f)$ then since $\sqrt{1-(\xi\vartheta_1^n)^2}\in\mathbb{R}\backslash\{0\}$, $\mathbb{F}(\mathbf{x};F)\approx1$ when
      \begin{equation}\label{FormulaeExp}
        k_-(\omega_f)\left(\xi\vartheta_1^n(z_1-x_1)-\sqrt{1-(\xi\vartheta_1^n)^2}(z_2-x_2)\right)=s\pi\quad\mbox{or}\quad s\pi+\frac{\pi}{2}
      \end{equation}
      for all $s\in\mathbb{Z}$, $f=1,2,\cdots,F$, and $n=1,2,\cdots,N$. Therefore, map of $\mathbb{F}(\mathbf{x};F)$ will offers exact location $\mathbf{z}$ but unexpected some ghost replicas will obstruct it.
  \item If $k_+(\omega_f)\ll k_-(\omega_f)$, (\ref{FormulaeExp}) becomes
      \begin{align*}
        &k_-(\omega_f)\left(\xi\vartheta_1^n(z_1-x_1)-\sqrt{1-(\xi\vartheta_1^n)^2}(z_2-x_2)\right)\\
        &=k_+(\omega_f)\vartheta_1^n(z_1-x_1)-\sqrt{k_-(\omega_f)^2-(k_+(\omega_f)\vartheta_1^n)^2}(z_2-x_2)\\
        &\approx k_+(\omega_f)\vartheta_1^n(z_1-x_1)-k_-(\omega_f)(z_2-x_2)=s\pi\quad\mbox{or}\quad s\pi+\frac{\pi}{2}
      \end{align*}
      for all $s\in\mathbb{Z}$, $f=1,2,\cdots,F$, and $n=1,2,\cdots,N$. This shows that $\mathbb{F}(\mathbf{x};F)\approx1$ if
      \[x_1=z_1,\quad x_2\approx z_2+\frac{s\pi}{k_-(\omega_f)}\quad\mbox{or}\quad x_1=z_1,\quad x_2\approx z_2+\frac{(2s+1)\pi}{2k_-(\omega_f)}.\]
      Since $k_-(\omega_f)$ is large enough, huge amounts of replicas will appear along the $y-$axis in the map of $\mathbb{F}(\mathbf{x};F)$ (see Figure \ref{Bothcontrast3}).
\end{enumerate}
The case of $\mathbb{F}_{\mathbf{D}}(\mathbf{x};F)$ can be handled in similar manner.

\subsection{Introduction to MUSIC and Kirchhoff migration}
By combining the results of \cite{AIL,PL2}, we can design a MUSIC-type image function at a single frequency $\omega$. We define a projection onto the null (or noise) subspace $\mathrm{P}_{\mathrm{\tiny noise}}$ as
\begin{equation}
\mathrm{P}_{\mathrm{\tiny noise}}(\mathbf{d}(\mathbf{x};\omega))=\sum_{m>M}\mathbf{U}_m(\omega)\mathbf{U}_m^*(\omega)\mathbf{d}(\mathbf{x};\omega).
\end{equation}
Then the image of $\mathbf{x}=\mathbf{z}_m$, $m=1,2,\cdots,M$, follows from the computation via a MUSIC-type imaging functional $\mathbb{F}_{\mathrm{\tiny MUSIC}}:\mathbb{R}_-^2\times\mathbb{R}\longrightarrow\mathbb{R}$,
\begin{equation}\label{ImagingMUSIC}
\mathbb{F}_{\mathrm{\tiny MUSIC}}(\mathbf{x};\omega)=\frac{1}{|\mathrm{P}_{\mathrm{\tiny noise}}(\mathbf{d}(\mathbf{x};\omega))|}.
\end{equation}
With this, we can find locations $\mathbf{x}=\mathbf{z}_m$ which satisfy $\mathbb{F}_{\mathrm{\tiny MUSIC}}(\mathbf{x};\omega)=\infty$.

We introduce the traditional Kirchhoff migration $\mathbb{F}_{\mathrm{KIR}}:\mathbb{R}_-^2\times\mathbb{R}\longrightarrow\mathbb{R}$
\begin{equation}\label{Kirchhoff}
  \mathbb{F}_{\mathrm{KIR}}(\mathbf{x};\omega)=|\mathbf{W_D}(\mathbf{x};\omega)^*\mathbb{K}\overline{\mathbf{W}}_{\mathbf{H}}(\mathbf{x};\omega)| =\left|\sum_{m=1}^{N}\sigma_m(\omega)\bigg(\mathbf{W_D}(\mathbf{x};\omega)^*\mathbf{U}_m(\omega)\bigg) \bigg(\mathbf{W_H}(\mathbf{x};\omega)^*\overline{\mathbf{V}}_m(\omega)\bigg)\right|
\end{equation}
Then similar to the filter function (\ref{Imagefunctionmultiple}), map of $\mathbb{F}_{\mathrm{KIR}}(\mathbf{x};\omega)$ will yields locations $\mathbf{z}_m$.

\subsection{Comparison of detection performance}\label{CIP}
Now, we will briefly compare the detection performance of (\ref{Imagefunctionmultiple}), (\ref{ImagingMUSIC}) and (\ref{Kirchhoff}). Roughly speaking, the following relationship holds for $F>1$
\begin{equation}\label{RelationImaging}
  \mathbb{F}_{\mathrm{\tiny MUSIC}}(\mathbf{x};\omega)\unlhd\mathbb{F}_{\mathrm{KIR}}(\mathbf{x};\omega)\unlhd\mathbb{F}(\mathbf{x};1)\unlhd\mathbb{F}(\mathbf{x};F),
\end{equation}
where $A\unlhd B$ means $B$ offers more accurate location $\mathbf{z}_m\in D_m$ than $A$.

First, based on the recent work \cite{AGKPS}, the relationship $\mathbb{F}_{\mathrm{\tiny MUSIC}}(\mathbf{x};\omega)\unlhd\mathbb{F}_{\mathrm{KIR}}(\mathbf{x};\omega)$ holds for homogeneous space case and we can easily verify that it also holds for two-layered medium problem interested herein (see \cite{P2} for numerical experiments).

Next, when $\mathbb{K}$ is affected by random noise, significant changes of singular values will appear. In this problem, such noise generates many nonzero singular values (see \cite[Figure 5.25]{AIL} so that (\ref{Kirchhoff}) generates a result with poor resolution. However, (\ref{Imagefunctionmultiple}) is not influenced by the singular values and therefore, map of $\mathbb{F}(\mathbf{x};1)$ yields a better result than $\mathbb{F}_{\mathrm{KIR}}(\mathbf{x};\omega)$, refer to Figure \ref{Compare}.

\begin{Rem}[Synthetic Aperture Radar (SAR) and Kirchhoff migration]
  Synthetic Aperture Radar (SAR) is one of the classical back-projection imaging technique developed for airborne radar applications (see \cite{BCP,CKI,TKSZAZDS,VMM} and references therein). Based on the research in \cite{TKSZAZDS}, results via SAR and Kirchhoff migration are almost the same.
\end{Rem}

Finally, based on the section \ref{Sec3sub}, we can examine the relationship $\mathbb{F}(\mathbf{x};F_1)\unlhd\mathbb{F}(\mathbf{x};F_2)$ holds for $F_1<F_2$. It is worth emphasizing that we can observe this relationship via statistical hypothesis testing (see \cite{AGKPS,FS,K} for detailed discussion).

\section{Numerical simulations and discussion}\label{Sec4}
\begin{figure}[!ht]
\begin{center}
\includegraphics[width=0.4\textwidth]{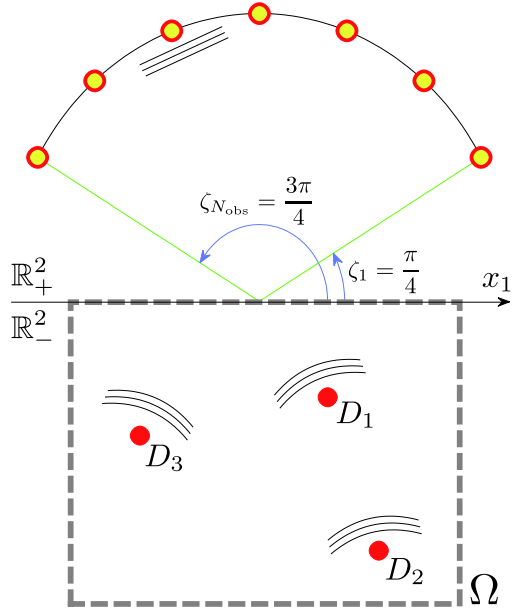}
\caption{\label{Conf}Sketch of the test configuration.}
\end{center}
\end{figure}

In this section, various numerical results are presented to demonstrate the effectiveness of the proposed algorithm. Same as the numerical configuration in \cite{AIL}, we choose three small homogeneous inhomogeneities, $D_1$, $D_2$ and $D_3$, embedded in the lower half-space. They are taken as ball of radius $r=0.1$ and are centered at $\mathbf{z}_1=(0.63,-2.47)$, $\mathbf{z}_2=(1.72,-4.97)$ and $\mathbf{z}_3=(-2,-3.63)$, respectively. The applied frequency is $\omega_f=\frac{2\pi}{\lambda_f}$, where $\lambda_f$, $f=1,2,\cdots,10$, are given wavelengths. In this paper, frequencies $\omega_f$ are equi-distributed within the interval $[2\pi,\frac{2\pi}{0.5}]$. The observation and incident directions $\boldsymbol\vartheta_j$ and $\boldsymbol\theta_l$ are taken as
\[\boldsymbol\vartheta_j=\left(\cos\zeta_j,\sin\zeta_j\right),\quad\zeta_j=\frac{\pi}{4}+\frac{(j-1)\pi}{2(N_{\mbox{\tiny obs}}-1)},
\quad\mbox{and}\quad
\boldsymbol\theta_l=-\left(\cos\varsigma_j,\sin\varsigma_j\right),\quad\varsigma_j=\frac{\pi}{4}+\frac{(l-1)\pi}{2(N_{\mbox{\tiny inc}}-1)},\]
respectively for $j=1,2,\cdots,N_{\mbox{\tiny obs}}$ and $l=1,2,\cdots,N_{\mbox{\tiny inc}}$. See Figure \ref{Conf} for an illustration of the test configuration.

Throughout this section, we adopt the squared search domain $\Omega=[-3,3]\times[-6,0]\subset\mathbb{R}_{-}^{2}$. The step size of the search points $\mathbf{x}\in\Omega$ is taken of the order of $0.05$ and vector $\mathbf{c}$ of (\ref{VecD}) is selected as
\begin{enumerate}
\item Permittivity contrast case: $\mathbf{c}=(1,0,0)$,
\item Permeability contrast case: $\mathbf{c}=(0,1,5)$,
\item Both permittivity and permittivity contrast case: $\mathbf{c}=(5,1,1)$.
\end{enumerate}
Note that this selection of $\mathbf{c}$ satisfies (\ref{SelectionC}). A detailed discussion about the choice of the vector $\mathbf{c}$ can be found in \cite[Section 4.2.1]{PL3}. In every example, the data set of the MSR matrix is computed within the framework of the Foldy-Lax equation, refer to \cite{DMG,PL2,TKDA}. Then, a white Gaussian noise with 20dB SNR(Signal to Noise Ratio) is added to the unperturbed data in order to show the robustness of the proposed algorithm. Note that from the various numerical experiments in \cite{AIL,P2,PL2}, similar results were obtained for both cases $\eps_+>\eps_-$ and $\eps_+=\eps_-$ (permeability and both contrast cases too). Thus, we do not consider the cases $\eps_+=\eps_-$ and/or $\mu_+=\mu_-$.

\subsection{Permittivity contrast case: $\eps_m\ne\eps_-$ and $\mu_m=\mu_-=\mu_+$}\label{sec4:1}
At this stage, we consider the purely dielectric contrast case. In this case, we set $\mu(\mathbf{x})=1$ for $\mathbf{x}\in\mathbb{R}^2$. For $\eps_+>\eps_-$ case, we choose the values $\eps_+=5$ and $\eps_-=4$ and permittivities $\eps_m$ of $D_m$ equal to $2,5,3$ for $m=1,2,3$. As already mentioned, the number of observation directions $N_{\mbox{\tiny obs}}$ must be smaller than the number of incidence directions $N_{\mbox{\tiny inc}}$. Hence, we choose $N_{\mbox{\tiny inc}}=10$ and $N_{\mbox{\tiny obs}}=6$ directions.

\begin{figure}[!ht]
\begin{center}
\includegraphics[width=0.49\columnwidth]{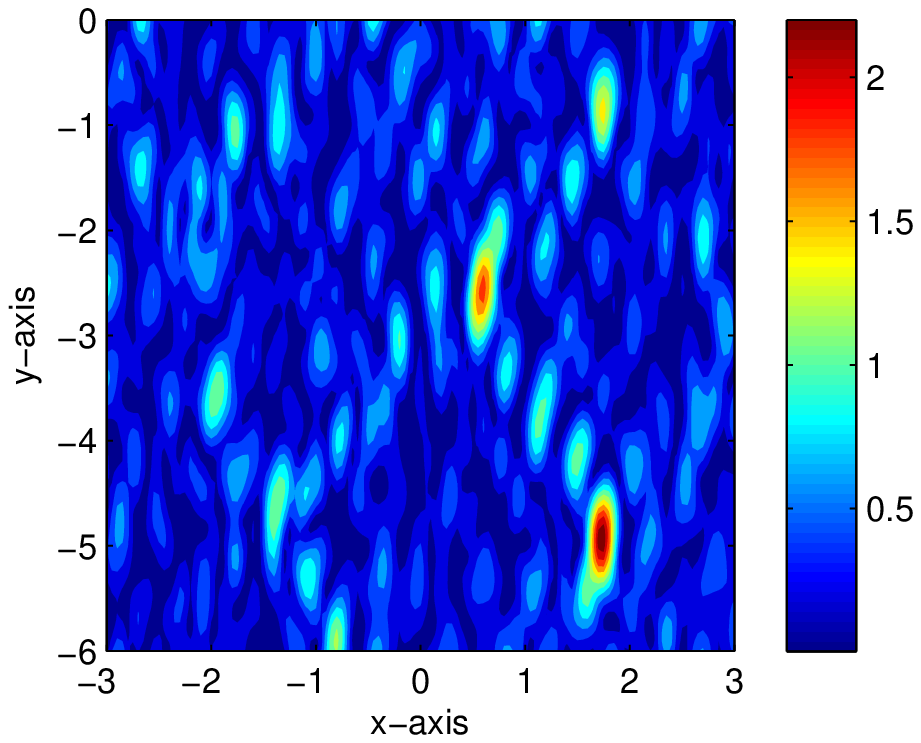}\label{Compare1}
\includegraphics[width=0.49\columnwidth]{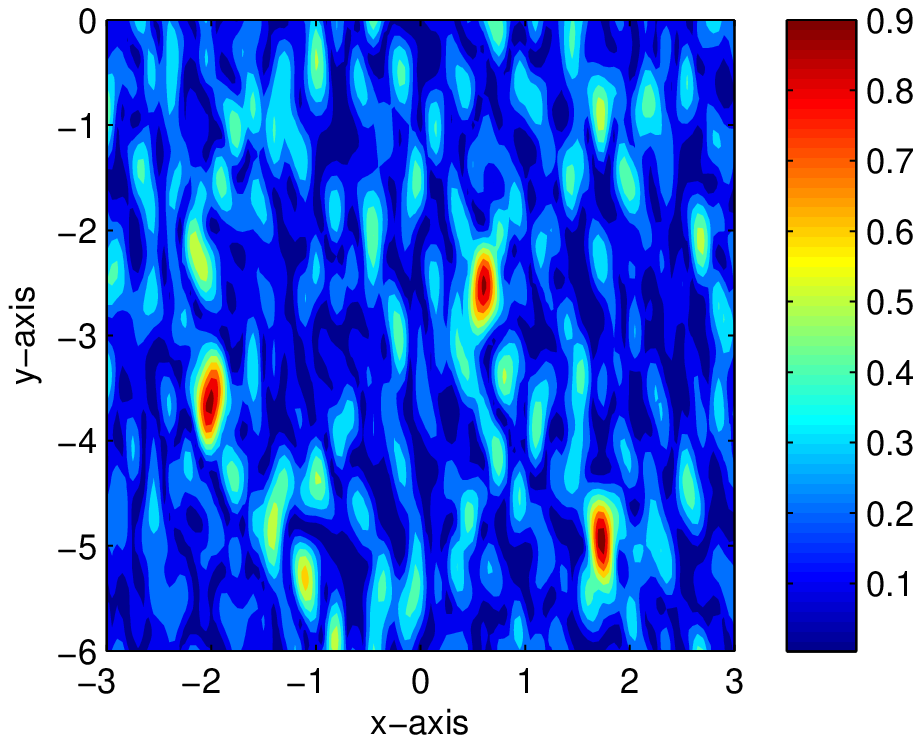}\label{Compare2}
\caption{(Compare the detection performance) Map of $\mathbb{F}_{\mathrm{KIR}}(\mathbf{x})$ (left) and $\mathbb{F}(\mathbf{x};1)$ (right) for $\omega=2\pi$ when $\eps_+=5$ and $\eps_-=4$.}\label{Compare}
\end{center}
\end{figure}

Let us check the detection performance of (\ref{Imagefunctionmultiple}) and (\ref{Kirchhoff}). As the results illustrated in Figure \ref{Compare}, map of $\mathbb{F}(\mathbf{x};1)$ offers more accurate location of small inhomogeneities (specially, $D_3$) than the one of $\mathbb{F}_{\mathrm{\tiny KIR}}(\mathbf{x};\omega)$. Moreover, by comparing the result in \cite[FIG 5.3]{AIL} and the map of $\mathbb{F}_{\mathrm{\tiny KIR}}(\mathbf{x};\omega)$, we can see that the relationship $\mathbb{F}_{\mathrm{\tiny MUSIC}}(\mathbf{x};\omega)\unlhd\mathbb{F}_{\mathrm{KIR}}(\mathbf{x};\omega)$ holds.

\begin{figure}[!ht]
\begin{center}
\includegraphics[width=0.49\columnwidth]{Frequency1.eps}
\includegraphics[width=0.49\columnwidth]{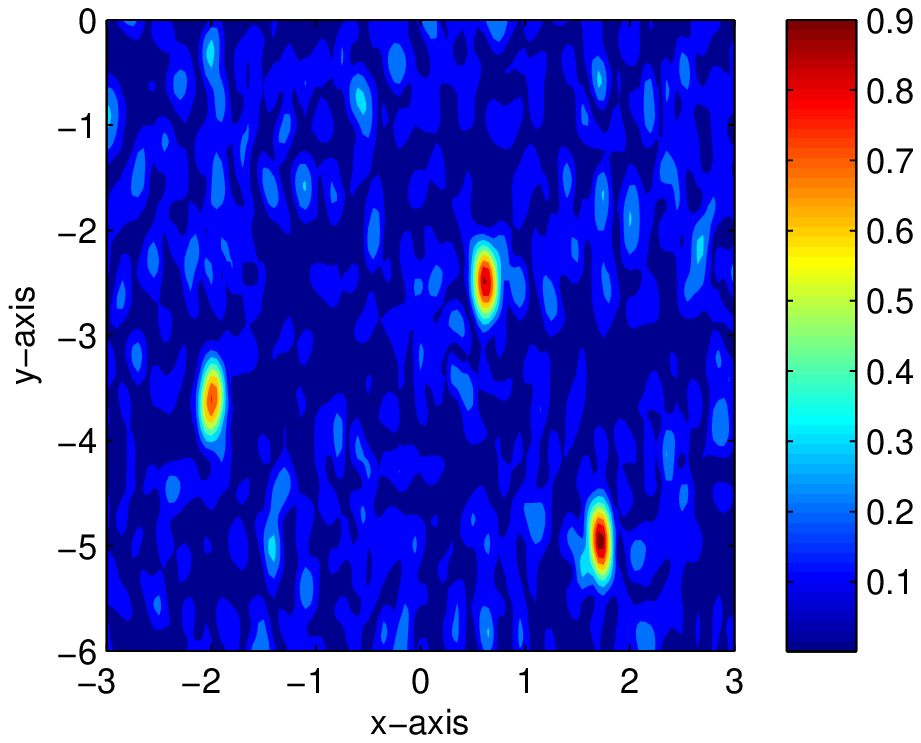}\\
\includegraphics[width=0.49\columnwidth]{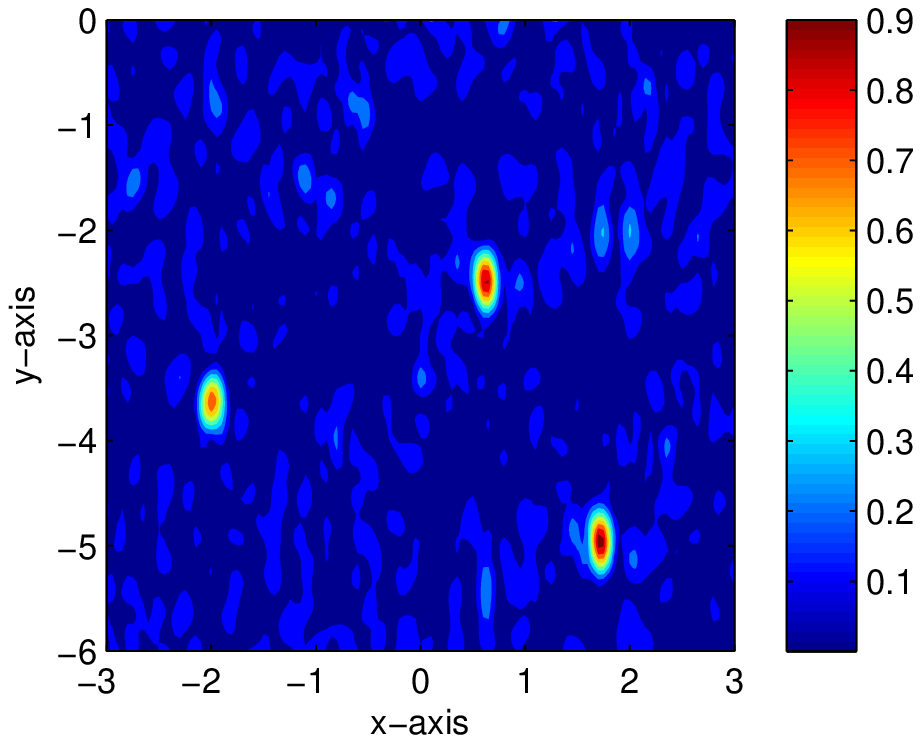}
\includegraphics[width=0.49\columnwidth]{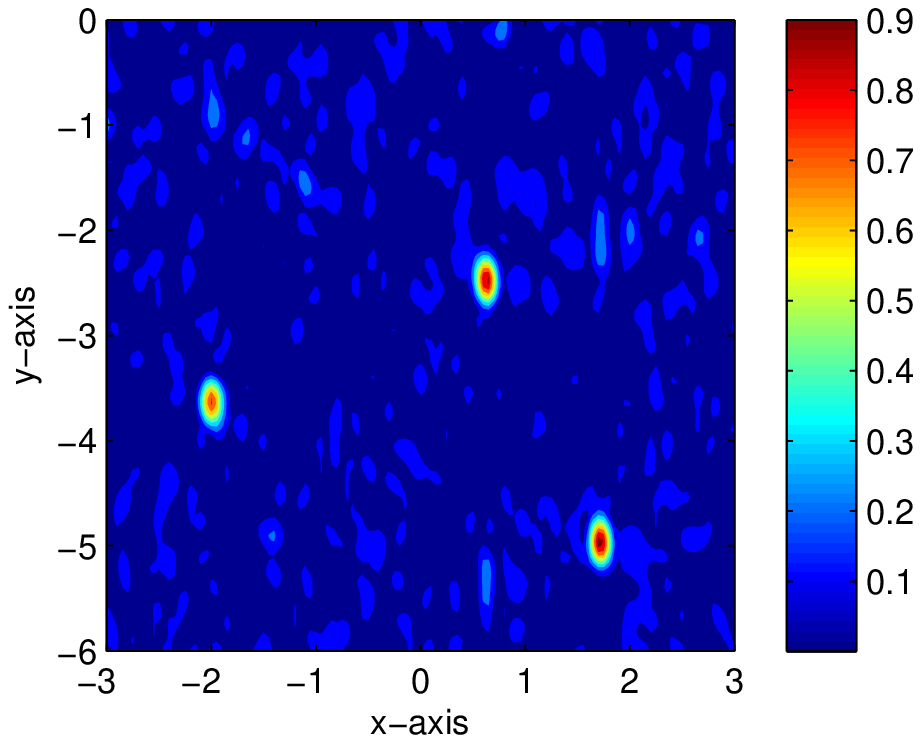}
\caption{(Influence of number of applied frequency) Map of $\mathbb{F}(\mathbf{x};F)$ for $F=1$ (top, left), $F=3$ (top, right), $F=5$ (bottom, left), and $F=7$ (bottom, right) when $\eps_+=5$ and $\eps_-=4$.}\label{Influence}
\end{center}
\end{figure}

Figure \ref{Influence} shows the influence of the number of applied frequencies. By comparing maps of $\mathbb{F}(\mathbf{x};F)$, we can easily observe that large number of $F$ guarantees an exact location of $D_m$. In various numerical tests, we observed that if one applied more than $F=7$ different frequencies, map of $\mathbb{F}(\mathbf{x};F)$ yields an accurate location of $D_m$, so we adopt $F=10$ frequencies in order to guarantee an admissible result.

\begin{figure}[!ht]
\begin{center}
\includegraphics[width=0.49\columnwidth]{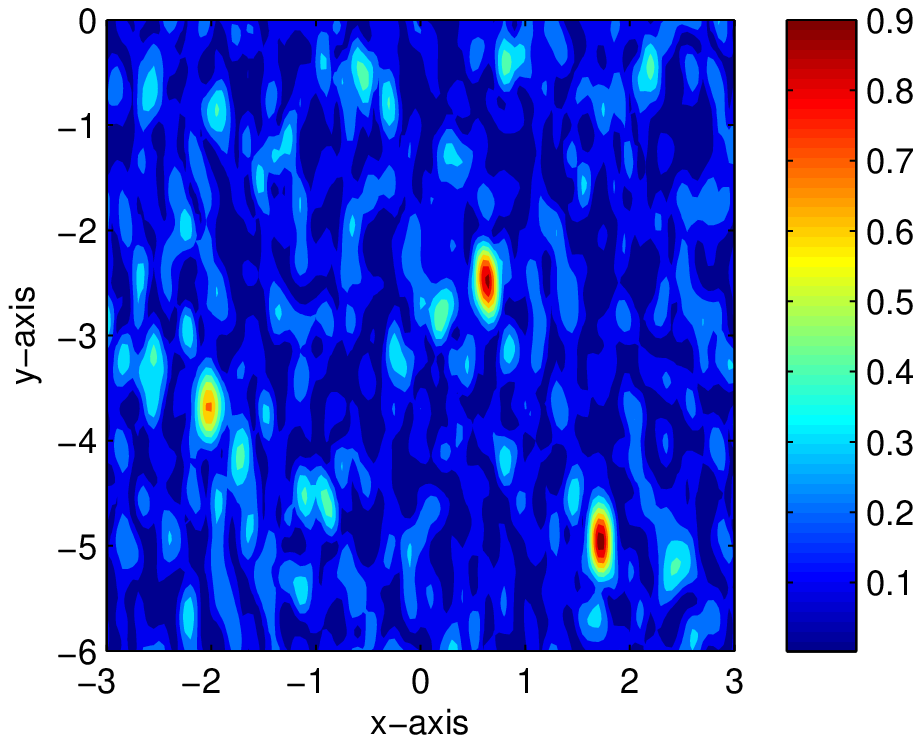}
\includegraphics[width=0.49\columnwidth]{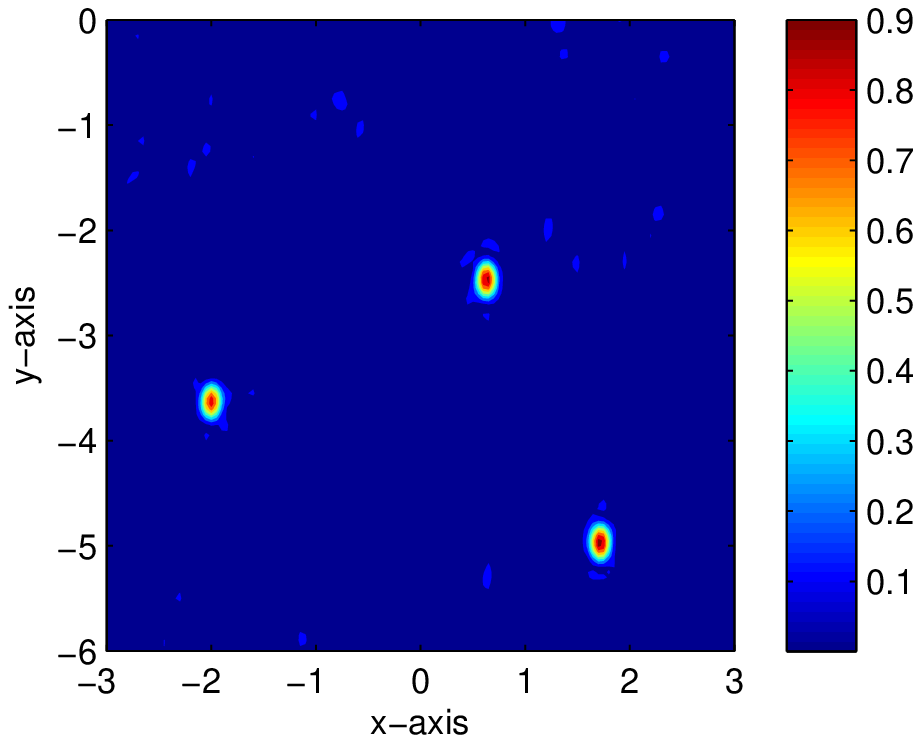}
\caption{(Influence of direction configuration) Map of $\mathbb{F}(\mathbf{x};F)$ for $N_{\mbox{\tiny obs}}=5$, $N_{\mbox{\tiny inc}}=7$, $F=5$ (left), and $N_{\mbox{\tiny obs}}=11$, $N_{\mbox{\tiny inc}}=13$, $F=10$ (right) when $\eps_+=5$ and $\eps_-=4$.}\label{Symmetric}
\end{center}
\end{figure}

In order to examine the influence of direction configuration, odd number of incident and observation directions is applied and corresponding results are illustrated in Figure \ref{Symmetric}. By comparing result in Figure \ref{Influence}, as we mentioned in section \ref{Sec3sub}, some replicas appeared under the odd number configuration but when the number of directions and frequencies is large enough, we can identify location of $D_m$ accurately.

\begin{figure}[!ht]
\begin{center}
\includegraphics[width=0.49\columnwidth]{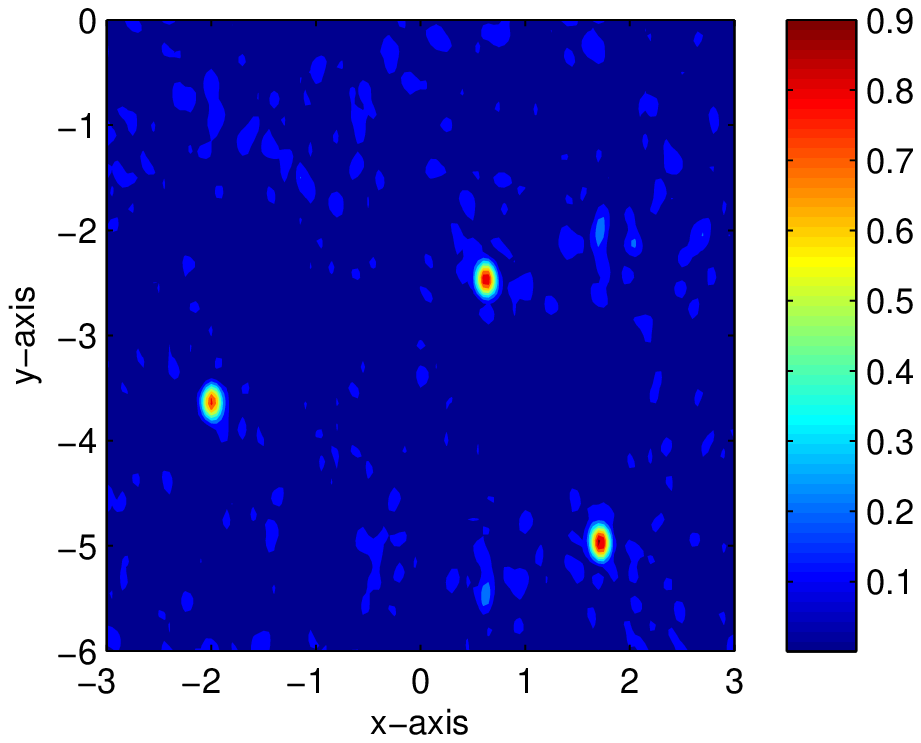}
\includegraphics[width=0.49\columnwidth]{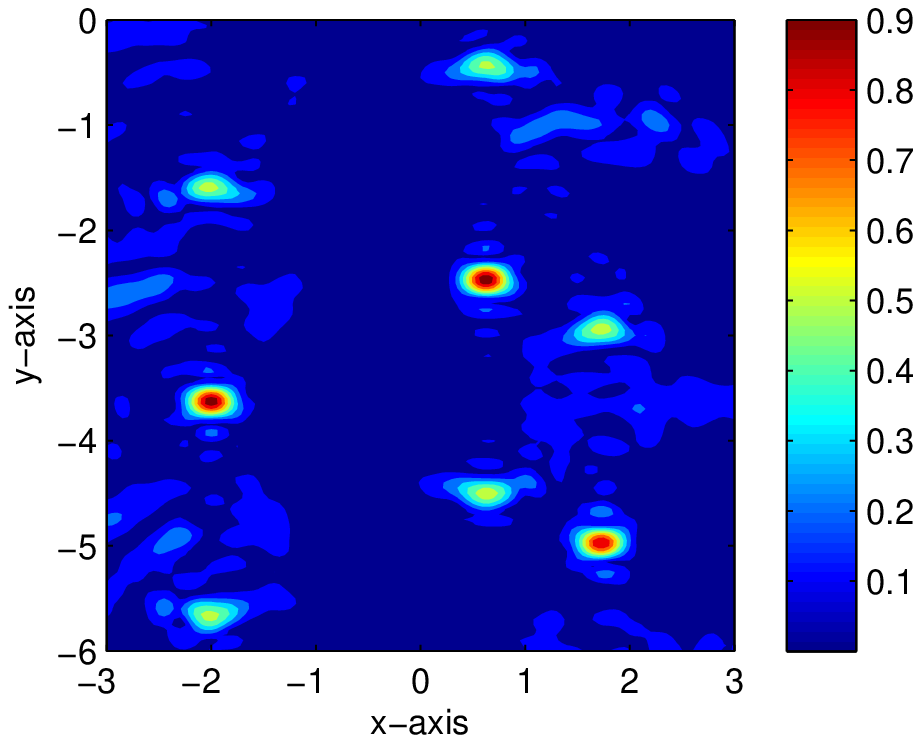}
\caption{(Permittivity contrast case) Map of $\mathbb{F}(\mathbf{x};10)$ when $\eps_+=5$, $\eps_-=4$ (left) and $\eps_+=1$, $\eps_-=5$ (right).}\label{Permittivitycontrast}
\end{center}
\end{figure}

For $\eps_+<\eps_-$ case, we choose the values $\eps_+=1$ and $\eps_-=5$ and permittivities $\eps_m$ of $D_m$ equal to $2,4,3$ for $m=1,2,3$. In this case, although a few ghost replicas appeared, the location of three inhomogeneities are successfully identified, refer to the right-hand side of Figure \ref{Permittivitycontrast}.

\subsection{Permeability contrast case: $\eps_m=\eps_-=\eps_+$ and $\mu_m\ne\mu_-$}\label{sec4:2}
\begin{figure}[!ht]
\begin{center}
\includegraphics[width=0.49\columnwidth]{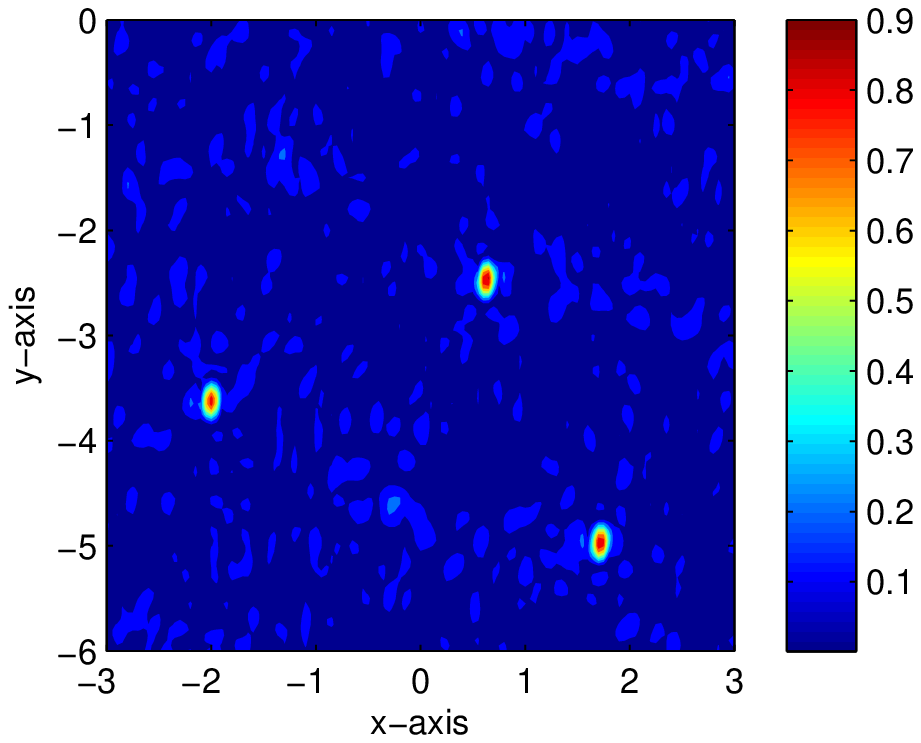}
\includegraphics[width=0.49\columnwidth]{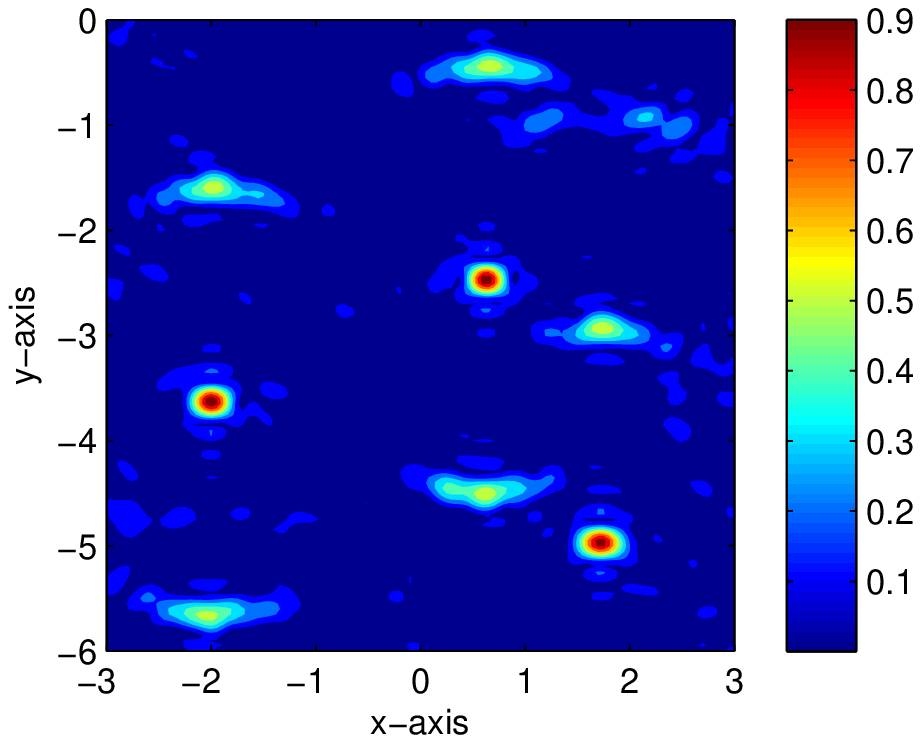}
\caption{(Permeability contrast case) Map of $\mathbb{F}(\mathbf{x};10)$ when $\mu_+=5$, $\mu_-=4$ (left) and $\mu_+=1$, $\mu_-=5$ (right)}\label{Permeabilitycontrast}
\end{center}
\end{figure}

Now, let us consider the purely magnetic permeability contrast case. In this case, we set $\eps(\mathbf{x})\equiv 1$ for $\mathbf{x}\in\mathbb{R}^2$. Similar to the section \ref{sec4:1}, we choose the values $\mu_+=5$ and $\mu_-=4$ and permeabilities $\mu_m$ of $D_m$ equal to $2,5,3$ for $m=1,2,3$. We apply $F=10$ frequencies, and $N_{\mbox{\tiny inc}}=14$ and $N_{\mbox{\tiny obs}}=8$ as directions of incidence and observation, respectively. For $\mu_+<\mu_-$ case, we choose the values $\mu_+=1$ and $\mu_-=5$ and permeabilities $\mu_m$ of $D_m$ equal to $2,4,3$ for $m=1,2,3$. Figure \ref{Permeabilitycontrast} shows the corresponding result. Similarly to Figure \ref{Permittivitycontrast}, the location of three inhomogeneities are successfully identified.

\subsection{Both permittivity and permeability contrast case: $\eps_m\ne\eps_-$ and $\mu_m\ne\mu_-$}\label{sec4:3}
\begin{figure}[!ht]
\begin{center}
\includegraphics[width=0.49\columnwidth]{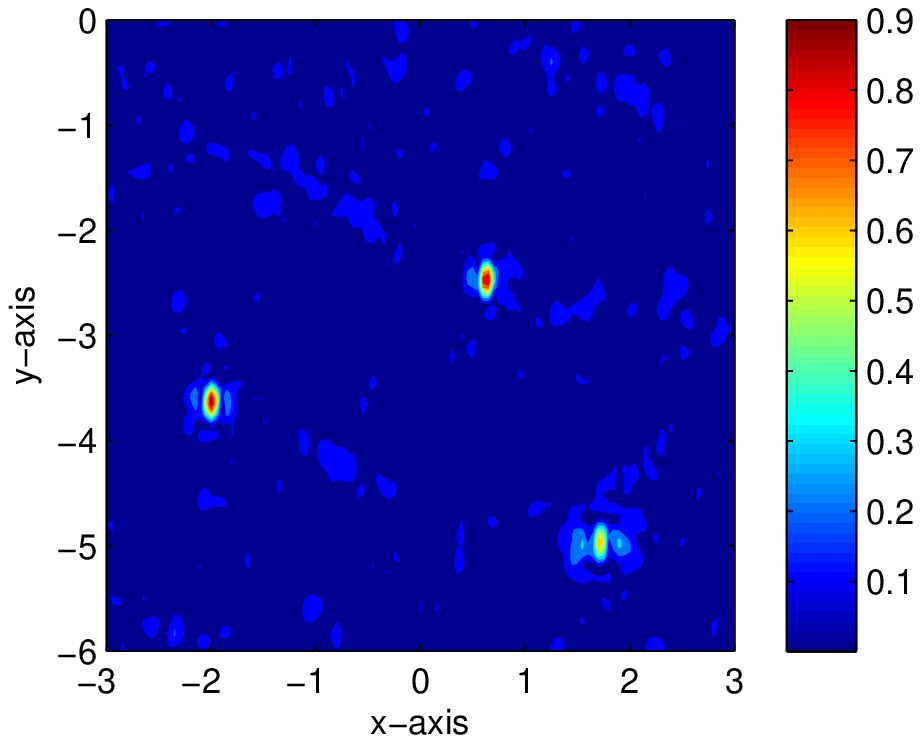}
\includegraphics[width=0.49\columnwidth]{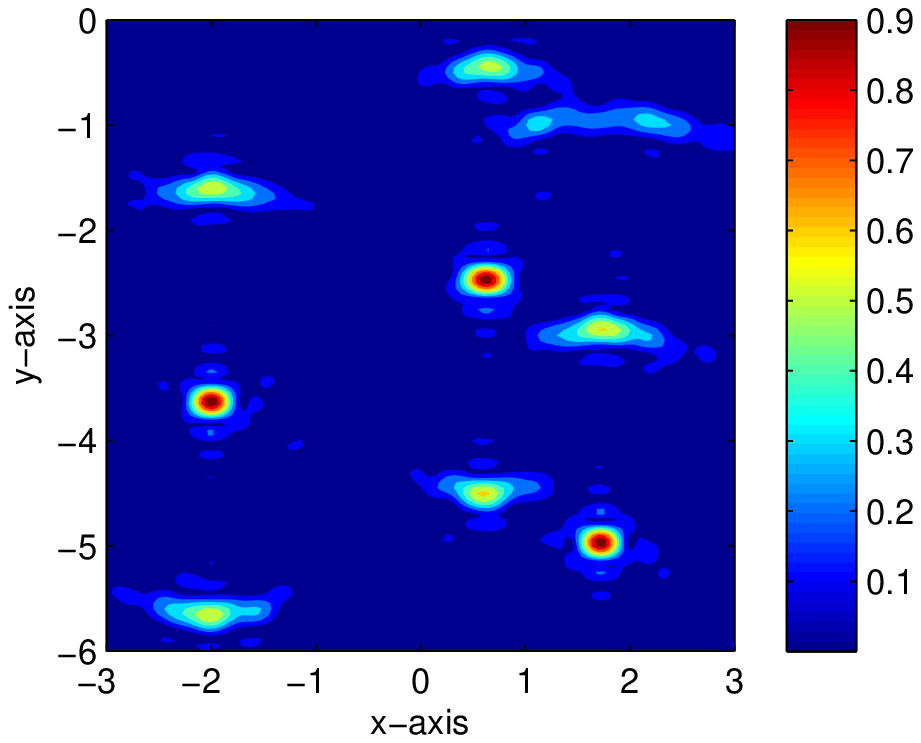}
\caption{(Both permittivity and permeability contrast case) Map of $\mathbb{F}(\mathbf{x};10)$ when $\eps_+=5$, $\eps_-=4$, $\mu(\mathbf{x})=1$ (left) and $\eps_+=1$, $\eps_-=5$, $\mu(\mathbf{x})=1$ (right).}\label{Bothcontrast1}
\end{center}
\end{figure}

\begin{figure}[!ht]
\begin{center}
\includegraphics[width=0.49\columnwidth]{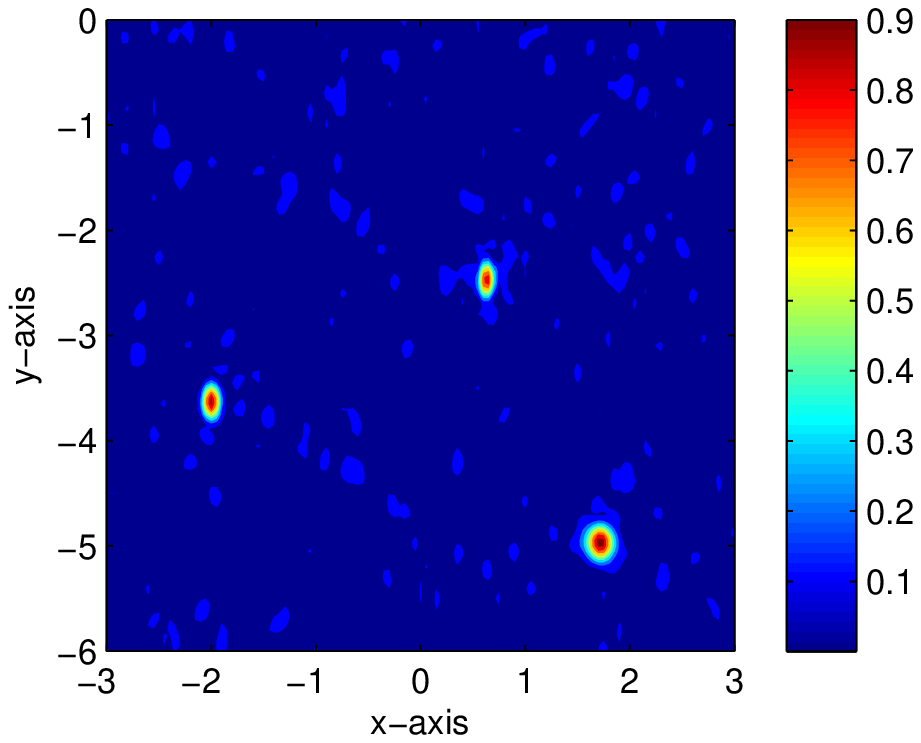}
\includegraphics[width=0.49\columnwidth]{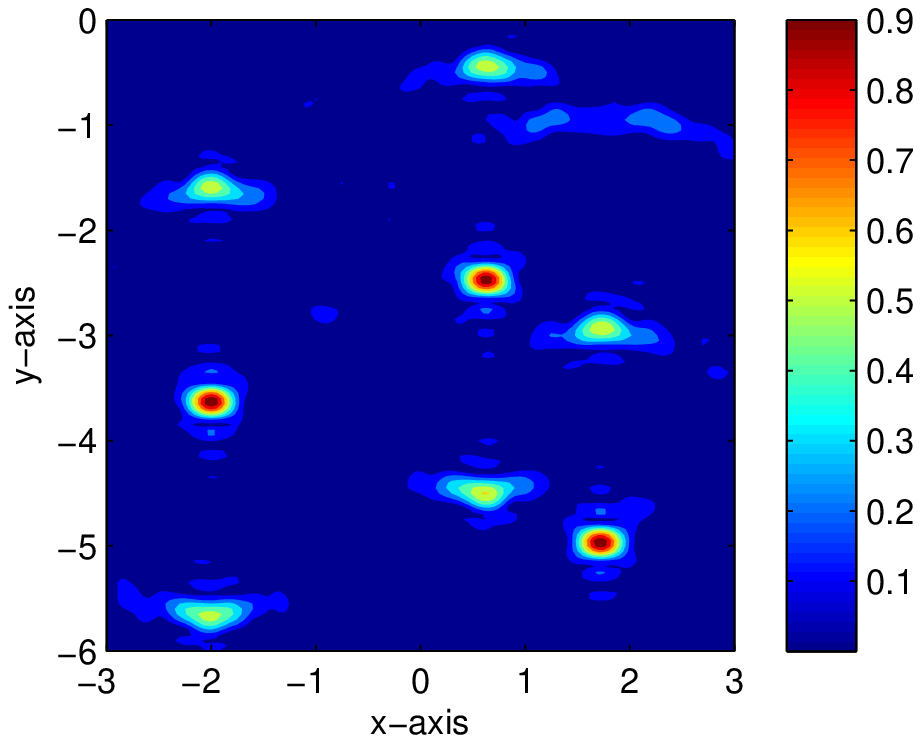}
\caption{(Both permittivity and permeability contrast case) Map of $\mathbb{F}(\mathbf{x};10)$ when $\eps(\mathbf{x})=1$, $\mu_+=5$, $\mu_-=4$ (left) and $\eps(\mathbf{x})=1$, $\mu_+=1$, $\mu_-=5$ (right).}\label{Bothcontrast2}
\end{center}
\end{figure}

In this case, we consider both permittivity and permeability contrast case. Three different situations of interest are considered:
\begin{itemize}
  \item $\eps_+\ne\eps_-$ and $\mu_+=\mu_-$,
  \item $\eps_+=\eps_-$ and $\mu_+\ne\mu_-$,
  \item $\eps_+\ne\eps_-$ and $\mu_+\ne\mu_-$.
\end{itemize}
First, let us consider the case $\eps_+>\eps_-$ and $\mu_+=\mu_-$. We set $\mu(\mathbf{x})\equiv 1$ for $\mathbf{x}\in\mathbb{R}^2$, $\eps_+=5$ and $\eps_-=4$ and permittivities $\eps_m$ and permeabilities $\mu_m$ of $D_m$ equal to $2,5,3$ for $m=1,2,3$. We apply $F=10$ frequencies and adopt $N_{\mbox{\tiny inc}}=20$ and $N_{\mbox{\tiny obs}}=12$ as directions of incidence and observation, respectively. The result, as illustrated in Figure \ref{Bothcontrast1}, remains very good and should be acceptable. Now, let us consider the case $\eps_+<\eps_-$ and $\mu_+=\mu_-$. Let $\eps_+=1$ and $\eps_-=5$ and permittivities $\eps_m$ and permeabilities $\mu_m$ of $D_m$ are equal to $2,4,3$ for $j=1,2,3$ while keeping remaining test configurations. Similarly to the previous example, a good result appeared, refer to Figure \ref{Bothcontrast1}. With a similar argument, we can obtain a good result when $\eps_+=\eps_-$ and $\mu_+\ne\mu_-$, refer to Figure \ref{Bothcontrast2}.

\begin{figure}[!ht]
\begin{center}
\includegraphics[width=0.49\columnwidth]{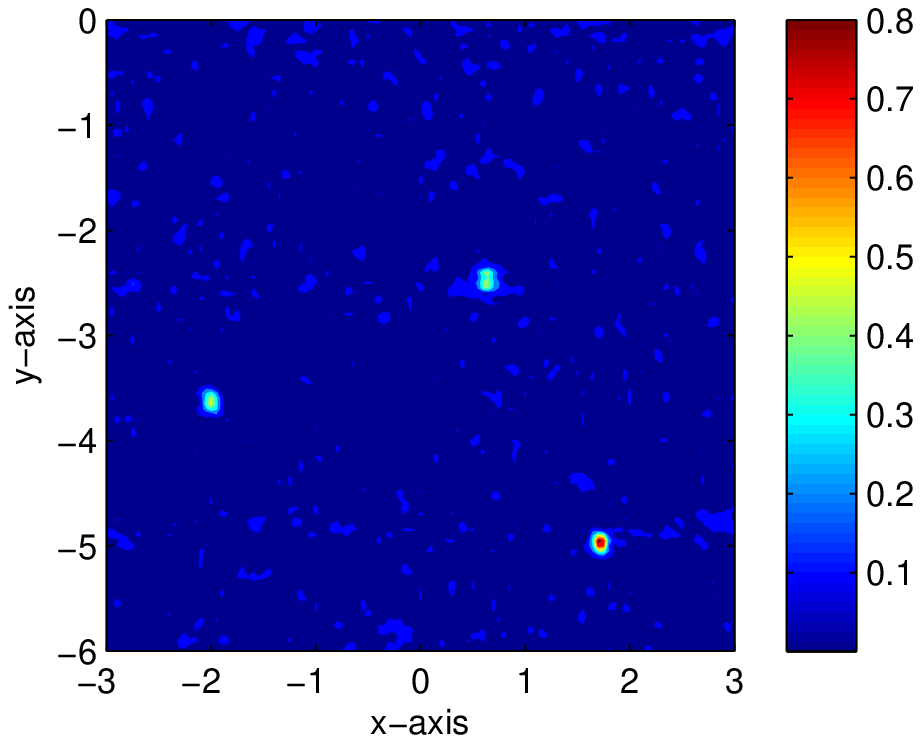}
\includegraphics[width=0.49\columnwidth]{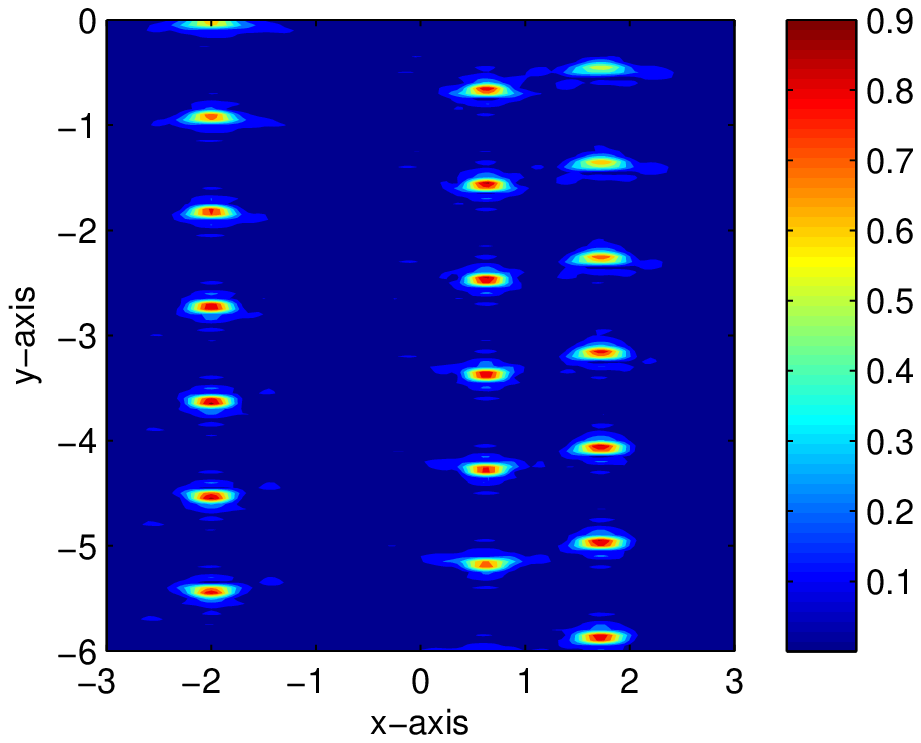}
\caption{(Both permittivity and permeability contrast case) Map of $\mathbb{F}(\mathbf{x};10)$ when $\eps_+=\mu_+=5$, $\eps_-=\mu_-=4$ (left) and $\eps_+=\mu_+=1$, $\eps_-=\mu_-=5$ (right)}\label{Bothcontrast3}
\end{center}
\end{figure}

For the case $\eps_+\ne\eps_-$ and $\mu_+\ne\mu_-$, we set $\eps_+=\mu_+=5$, $\eps_-=\mu_-=4$ and $\eps_+=\mu_+=1$, $\eps_-=\mu_-=5$ while keeping the configuration of the previous situation. The result is exhibited in Figure \ref{Bothcontrast3}. Although we could obtain a reasonably good result when $\eps_+>\eps_-$ and $\mu_+>\mu_-$, one can not determine accurate locations of inhomogeneities when $\eps_+<\eps_-$ and $\mu_+<\mu_-$ based on the discussions in section \ref{Sec3sub}.

Various results in this paper show that the proposed algorithm is very stable and effective but it still contains some factors for further improvements.

\subsection{Robustness with respect to random noise and influence of range of incident and observation direction}
\begin{figure}[!ht]
\begin{center}
\includegraphics[width=0.49\columnwidth]{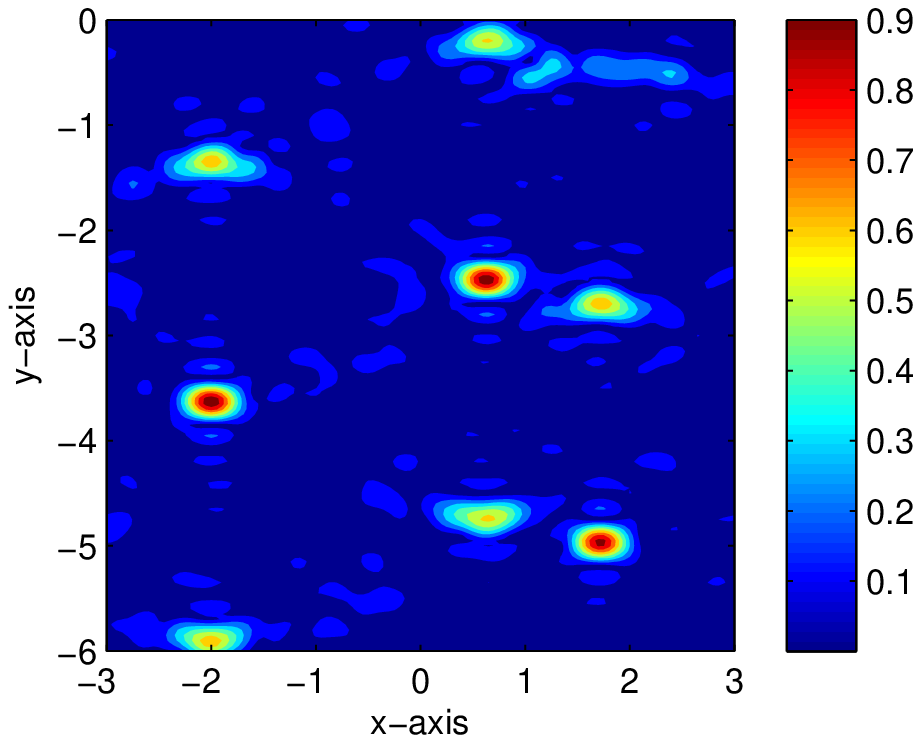}
\includegraphics[width=0.49\columnwidth]{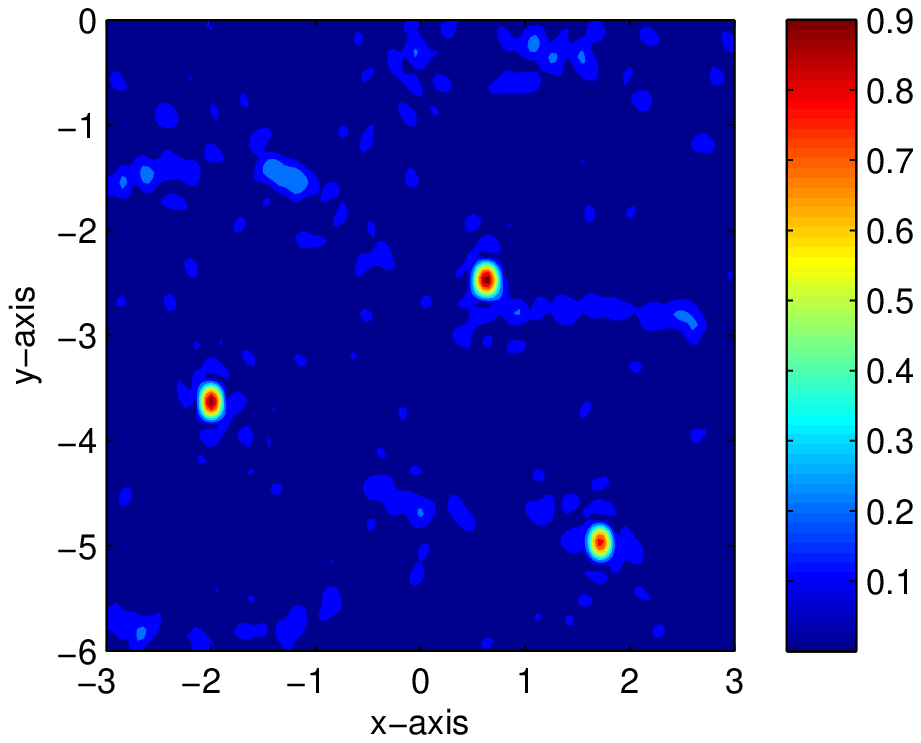}\\
\includegraphics[width=0.49\columnwidth]{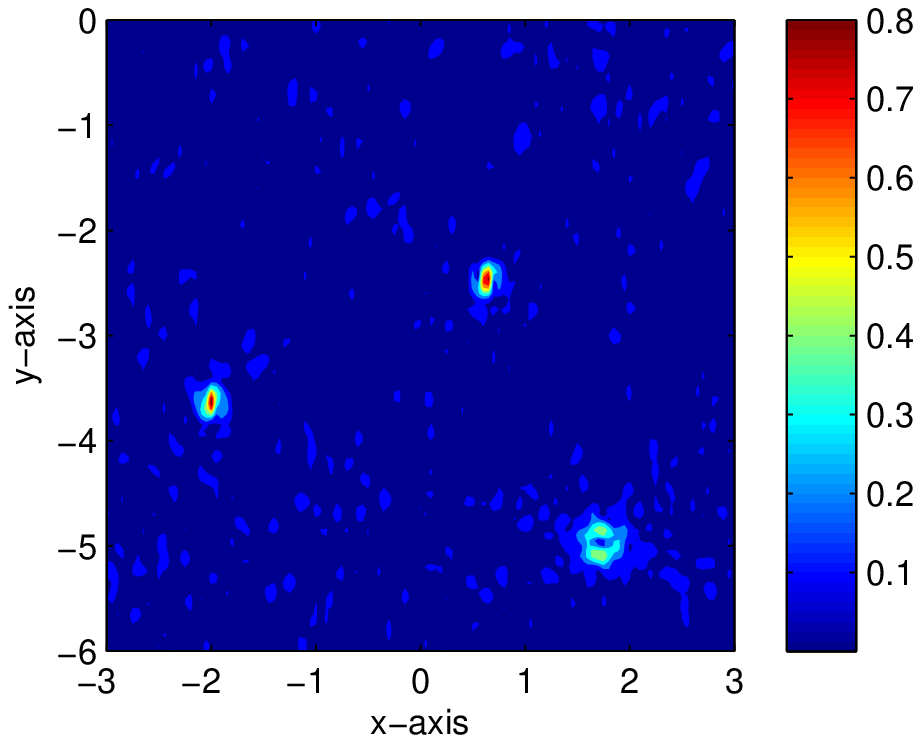}
\includegraphics[width=0.49\columnwidth]{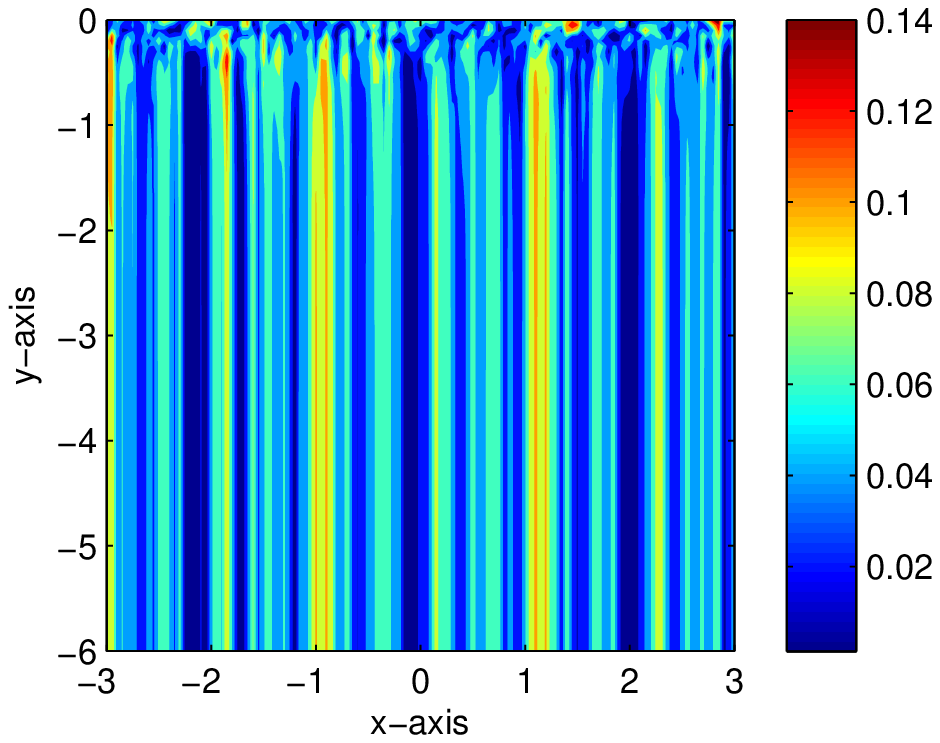}
\caption{(Both permittivity and permeability contrast case) Map of $\mathbb{F}(\mathbf{x};10)$ when $\frac{5\pi}{12}\leq\zeta_j,\varsigma_l\leq\frac{7\pi}{12}$ (top, left), $\frac{\pi}{3}\leq\zeta_j,\varsigma_l\leq\frac{2\pi}{3}$ (top, right), $\frac{\pi}{6}\leq\zeta_j,\varsigma_l\leq\frac{5\pi}{6}$ (bottom, left), and $\frac{\pi}{12}\leq\zeta_j,\varsigma_l\leq\frac{11\pi}{12}$ (bottom, right)}\label{InfluenceRange}
\end{center}
\end{figure}

At this moment, we add a white Gaussian noise with 10dB SNR to the unperturbed data and change the range of incident and observation directions. For this purpose, we adopt the same test configuration as section \ref{sec4:3} except that the ranges of observation directions
\[\boldsymbol\vartheta_j=-\left(\cos\zeta_j,\sin\zeta_j\right),\quad j=1,2,\cdots,N_{\mbox{\tiny obs}},\]
are varied by changing the values of $\zeta_1$ and $\zeta_{N_{\mbox{\tiny obs}}}$. Incident directions are also changed similarly. Figure \ref{InfluenceRange} shows the map of $\mathbb{F}(\mathbf{x};10)$ when $\eps_+=\mu_+=5$, $\eps_-=\mu_-=4$ under the narrow and wide range of observation and incident direction configurations. Regarding the top, left-hand side of Figure \ref{InfluenceRange}, we can see that under the narrow range of incident and observation direction configuration, map of $\mathbb{F}(\mathbf{x};10)$ contains some ghost replicas similar to the case of $\eps_+<\eps_-$ (see Figure \ref{Bothcontrast1}). From the top, right-hand side of Figure \ref{InfluenceRange}, we can clearly identify three locations of $\mathbf{z}_m$. Therefore, we can conclude that the proposed algorithm is robust with respect to the large amount of random noise since the normalized filter function (\ref{Imagefunctionmultiple}) is not significantly influenced by the noise. Unfortunately, based on the section \ref{Sec3sub}, by regarding the bottom line of Figure \ref{InfluenceRange}, map of $\mathbb{F}(\mathbf{x};F)$ yields poor result when the range of directions become wider. Hence, we can conclude that when the range of incident and observation directions become wider, we cannot find the location $\mathbf{z}_m$ via map of $\mathbb{F}(\mathbf{x};F)$. This shows a limitation of the proposed algorithm.

\subsection{On the Rayleigh resolution limit}
\begin{figure}[!ht]
\begin{center}
\includegraphics[width=0.49\columnwidth]{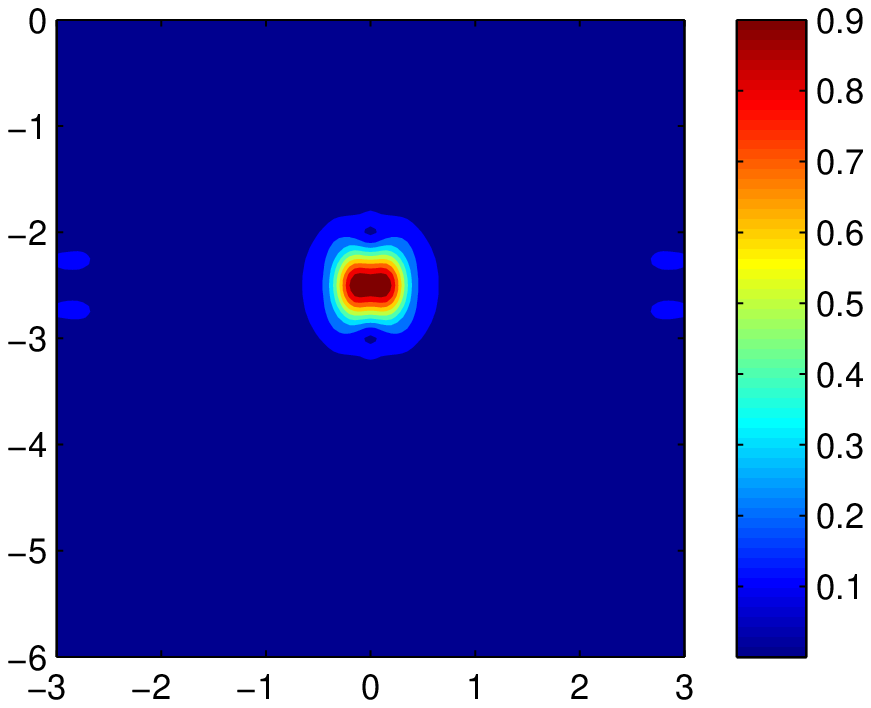}
\includegraphics[width=0.49\columnwidth]{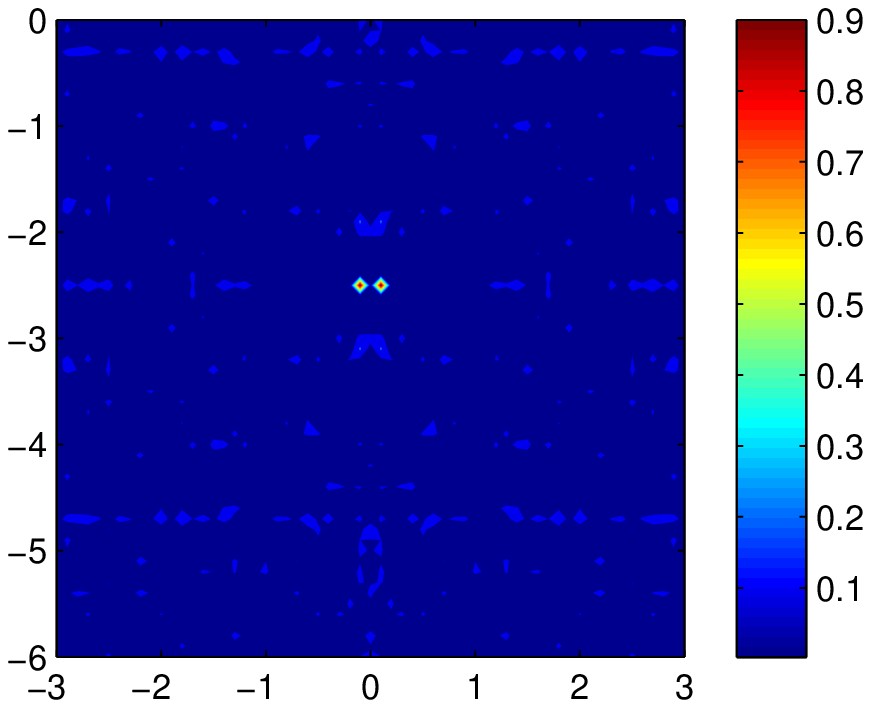}
\caption{(Permittivity contrast case with $\eps_+=5$ and $\eps_-=4$) Map of $\mathbb{F}(\mathbf{x};10)$ when $\lambda_1=2$, $\lambda_{10}=1$ (left) and $\lambda_1=0.2$, $\lambda_{10}=0.1$ (right).}\label{Rayleigh}
\end{center}
\end{figure}

Now, we briefly consider the image resolution. From the Rayleigh resolution limit, we can distinguish any inhomogeneities $D$ and $D'$ when
\[\mbox{dist}(D,D')\geq\frac{\lambda}{2}.\]
In order to examine such phenomenon, we consider the detected location of two disks $D_4$ and $D_5$ of the same radius $r=0.01$ with permittivity contrast case. The centers of $D_4$ and $D_5$ are selected as $\mathbf{z}_4=(-0.1,-2.5)$ and $\mathbf{z}_5=(0.1,-2.5)$, respectively. In Figure \ref{Rayleigh}, we illustrate corresponding results with large and small wavelengths while keeping remaining test configurations. Therefore, it is hard to distinguish two inhomogeneities from the image via map of $\mathbb{F}(\mathbf{x};10)$ with large wavelengths. On the other hand, when we apply smaller wavelengths (i.e., higher frequency) than the previous one, the two inhomogeneities become distinguishable.

\section{Concluding remarks}\label{Sec5}
In this paper, we suggest a location search algorithm operated at several time-harmonic frequencies in order to find accurate locations of small electromagnetic inhomogeneities completely embedded within a homogeneous lower half-space. The approach is based on the asymptotic formulation due to the existence of small electromagnetic inhomogeneities. Throughout numerical simulations, we can conclude that the proposed algorithm not only performs quite well even in the existence of random noise but also successfully improves existing limitations (poor longitudinal resolution against an excellent transverse resolution) of MUSIC algorithm proposed in \cite{AIL} and Kirchhoff migration. In addition, it still has some points of improvement for finding the locations of inhomogeneities under the situation $\eps_+<\eps_-$, $\mu_+<\mu_-$ and wide range of incident/observation directions.

It is worth mentioning that such results obtained at low computational costs can be a good initial guess of a level-set evolution \cite{ADIM,DEKPS,DL,PL4} or of any other standard iterative algorithm. Although only two-dimensional problem have been considered herein, we expect that the proposed strategy, e.g., asymptotic formula, filter design, etc., could be extended to the three-dimensional problem, refer to \cite{AILP,IGLP} for related works. Moreover, inconveniences of the proposed algorithm are also the same as those of SAR\footnote{This was suggested to the authors by one of the anonymous referees}. Hence, comparison of SAR and the proposed algorithm will be an interesting subject.

Finally, we would like to emphasize that the proposed algorithm can be extended to the shape identification of electromagnetically thin, arc-like, penetrable inhomogeneities, refer to \cite{P1,P2,PL2}. Although, further mathematical investigation is necessary, we believe that it can also be extended to the identification of small or extended perfectly conducting cracks.

\section{Acknowledgement}
We would like to acknowledge two anonymous referees for their precious comments. Won-Kwang Park was supported by the Basic Science Research Program through the National Research Foundation of Korea (NRF) funded by the Ministry of Education, Science and Technology (No. 2012-0003207), the WCU(World Class University) program through the National Research Foundation of Korea(NRF) funded by the Ministry of Education, Science and Technology R31-10049, and the research program of Kookmin University in Korea. Taehoon Park was supported by the research program of Kookmin University in Korea.

\end{document}